\documentclass{egarxiv}
\ConferencePaper   % uncomment for Conference submission
\WsPaper

\usepackage[T1]{fontenc}
\usepackage{dfadobe}  

\usepackage{cite}  % comment out for biblatex with backend=biber
\BibtexOrBiblatex
\electronicVersion
\PrintedOrElectronic
\ifpdf \usepackage[pdftex]{graphicx} \pdfcompresslevel=9
\else \usepackage[dvips]{graphicx} \fi

\usepackage{cite}  % comment out for biblatex with backend=biber

\usepackage{booktabs} % For formal tables
\usepackage{adjustbox}
\usepackage{amsfonts}
\usepackage{xspace}
\usepackage{relsize}
\def\code#1{{{\relsize{-1}\texttt{#1}}}\xspace}

\usepackage{color}

\definecolor{dkgreen}{rgb}{0,0.6,0}
\definecolor{gray}{rgb}{0.5,0.5,0.5}
\definecolor{mauve}{rgb}{0.58,0,0.82}

\definecolor{dkgreen}{rgb}{0,0.6,0}
\definecolor{dkblue}{rgb}{0,0,0.6}
\definecolor{gray}{rgb}{0.5,0.5,0.5}
\definecolor{mauve}{rgb}{0.58,0,0.82}

\usepackage{xcolor}
\definecolor{commentgreen}{RGB}{2,112,10}
\definecolor{eminence}{RGB}{108,48,130}
\definecolor{weborange}{RGB}{255,165,0}
\definecolor{frenchplum}{RGB}{129,20,83}

\usepackage{listings}
\title[AMR Iso-Surface Extraction]{A Simple, General, and Parallel Method for
  Extracting Crack-free
  Iso-Surfaces from
  Adaptive Mesh Refinement (AMR) Data\\[-1.1em]}

\title[AMR Iso-Surface Extraction]{A Simple, General, and GPU Friendly Method for
  Computing Dual Mesh and Iso-Surfaces of 
  Adaptive Mesh Refinement (AMR) Data\\[-1.1em]}

\author[I. Wald]{
  \parbox{\textwidth}{\centering I.~Wald
    \orcid{0000-0003-0046-713X}
  }
  \\
    {\parbox{\textwidth}
      {\centering %$^1$
        NVIDIA
      }
    }
}

\author{Ingo Wald}
\begin{document}
%\begin{teaserfigure}
\teaser{
  \vspace*{-9mm}
  \centering
  \resizebox{0.97\textwidth}{!}{
    a)
    \includegraphics[height=4cm]{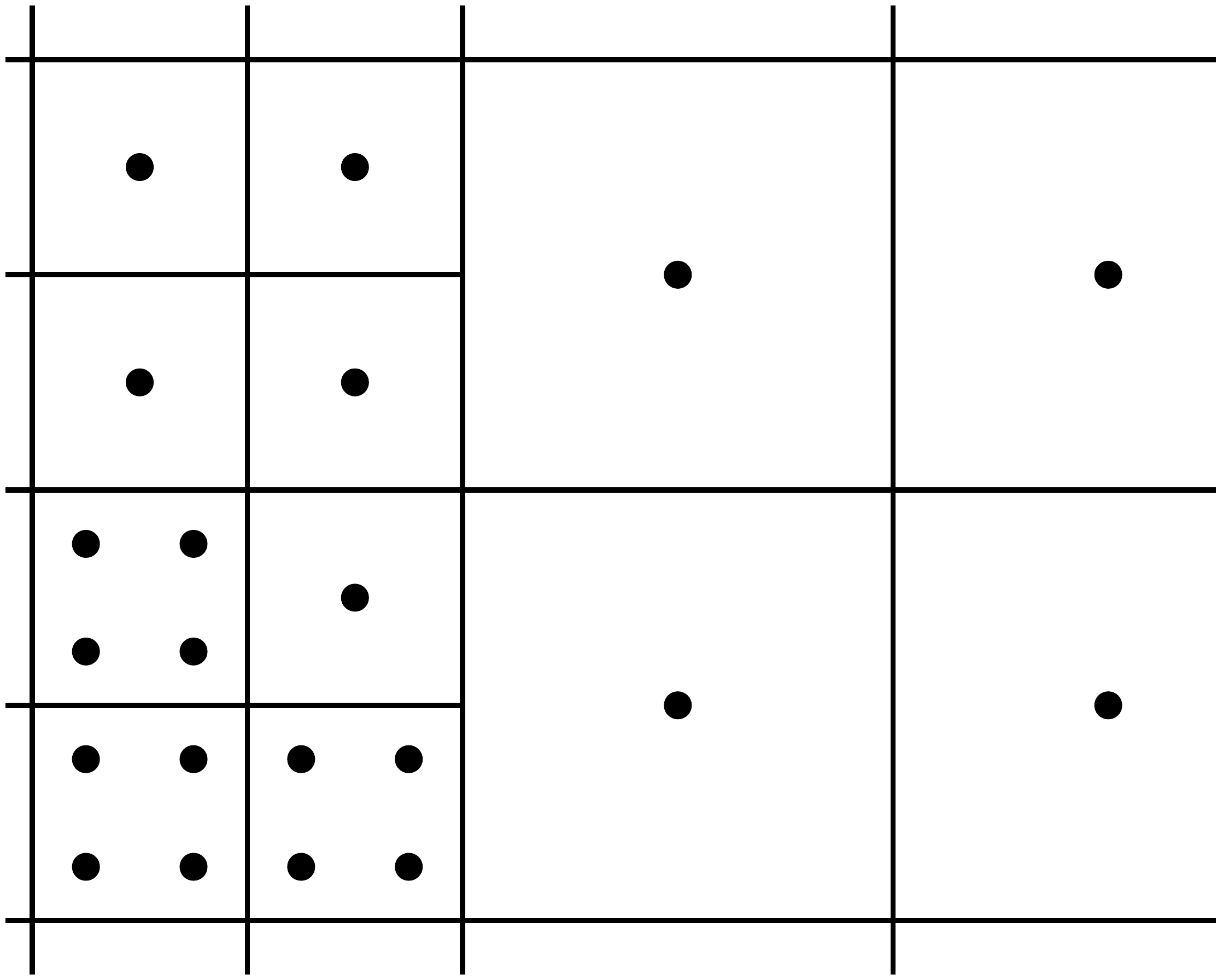}
    \quad
    \quad
    b)
    \includegraphics[height=4cm]{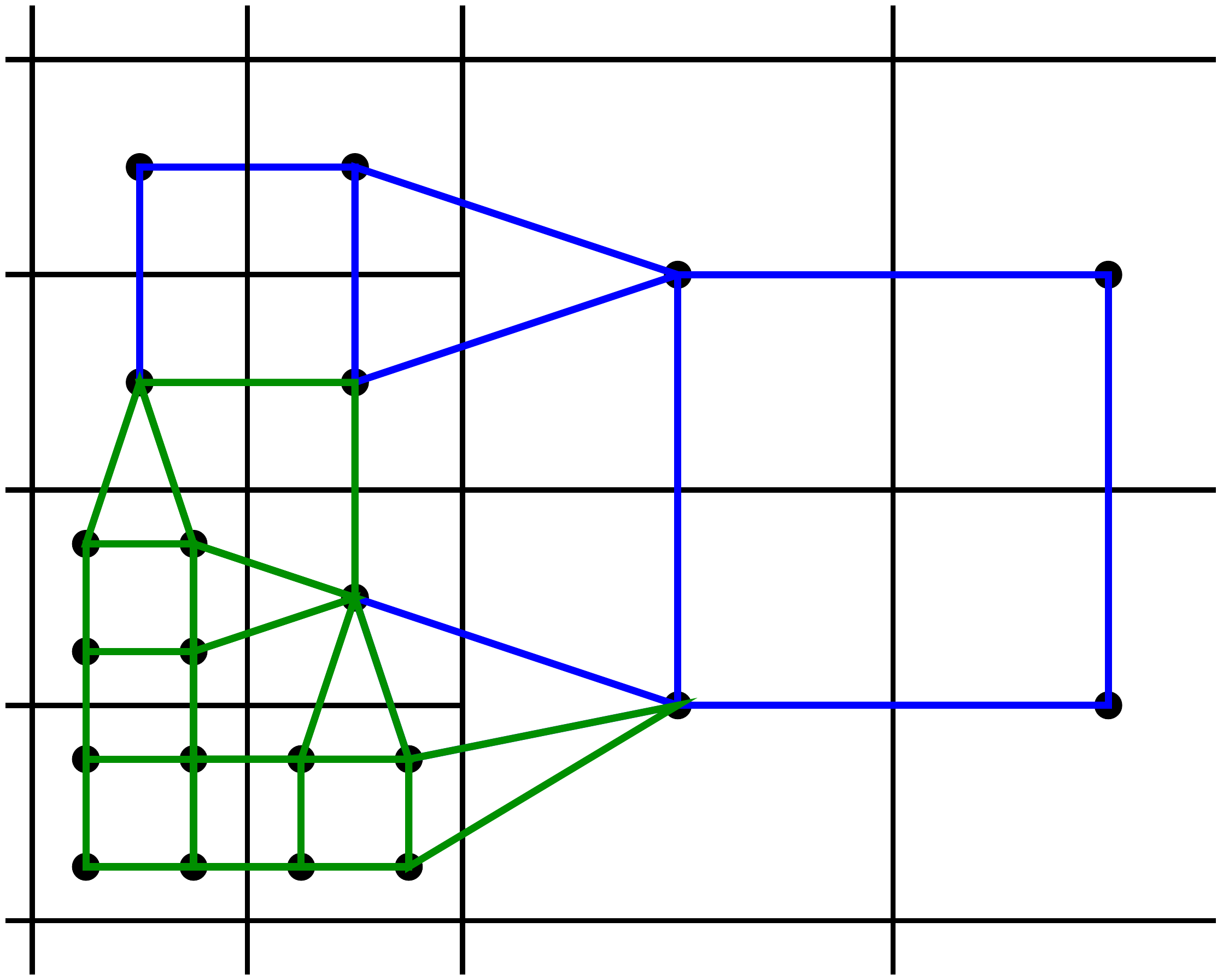}
    \quad
    \quad
    c)
    \includegraphics[height=4cm]{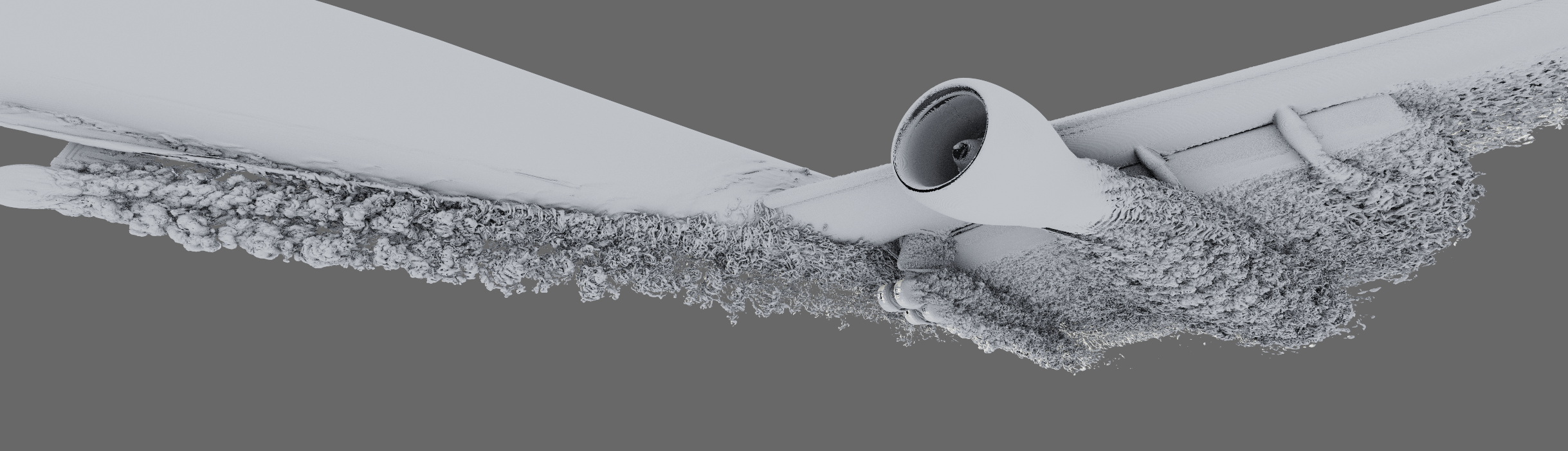}
  }\vspace{-.8em}
  \caption{\label{fig:teaser} a) Given an arbitrary set of invidiual
    AMR cells (that could come from either block-structured or octree
    AMR data) out method provides an easy, programmatically simple,
    and trivially parallel way of computing all cells of the dual mesh
    (b), which can then, for example be fed into the Marching Cubes
    case tables to produce a crack-free polygonal iso-surface.  c) A
    130 million triangle iso-surface ($||vorticity||$ field,
    $\rho=1000$) of the NASA Exajet model, which consists of $626M$
    individual AMR cells (with holes, and without any apparent
    hierarchy). Computing this surface with a CUDA implementation of
    our method took 8.4 seconds on a NVIDIA RTX~8000 GPU.  } }
%\end{teaserfigure}
%

\maketitle

%\abstract{
\begin{abstract}
  We propose a novel approach to extracting crack-free
  iso-surfaces from Structured AMR data that is more general than
  previous techniques, is trivially simple to implement, requires no
  information other than the list of AMR cells, and works, in
  particular, for different AMR formats including octree AMR, block-structured
  AMR with arbitrary level differences at level boundaries,
  and AMR data that consist of individual
  cells without any existing grid structure. We describe both the
  technique itself and a CUDA-based GPU implementation of this
  technique, and evaluate it on several non-trivial AMR data sets.
\end{abstract}
  %}

%\keywords{Ray Tracing, Adaptive Mesh Refinement (AMR)}
%\begin{document}
%\firstsection{Introduction}

\parskip 4.5pt plus 1pt minus 1.5pt

\section{Introduction}

Adaptive Mesh Refinement (AMR) was first introduced by Berger et
al.~\cite{berger1984adaptive,berger1989local}, and has since become
one of the most widely used methods for scientific simulation codes.
Its base idea is that a initially coarse domain subdivision gets
adaptively refined based where the simulation has the most need
for finer discretization---which significantly reduces the number of
discretization elements (and thus, memory storage) required to reach a
certain accuracy.

Visualizing such data often requires the computation of a polygonal
iso-surface over this data; this can then be rendered, typically
color-mapped with one or more additional attributes. Unfortunately,
computing crack-free iso-surfaces can be challenging due to the
irregular nature of the AMR cells; in particular, the dual mesh of a
structured AMR data set is a \emph{un}structured mesh that in the
general case can contain unstructured elements ranging from tetrahedra
to non-rectangular hexahedra with curved sides.

%% Some progress on defining such interpolants has recently been made by
%% Wald et al~\cite{wald:17:AMR} and Wang et al.~\cite{wang:18:iso-amr};
%% however, the default method for visualizing AMR data is still to
%% extract an explicit polygonal iso-surface, which may also have
%% applications beyond rendering. For extracting a crack-free
%% iso-surface, the state of the art today is to first compute the
%% (unstructured) dual mesh of the AMR data by ``stitching'' level
%% boundaries with unstructured-mesh elements such as tetrahedra,
%% pyramids, wedges, and general hexahedra.

Today, the best known method for computing crack-free iso-surfaces of
AMR data is what we call the ``stitching'' method by Weber et
al.~\cite{weber2003extraction,weber2012efficient}: this method walks the
brick boundaries, and stitches these with unstructured elements that are picked
by using pre-computed case tables. This method is widely used,
but  relies on certain
assumptions---in particular, that the input is block-structured AMR
with no more than one level differences at block boundaries---that are
not always met even for block-structured AMR, and not at all for
octree-AMR. Furthermore, the non-trivial case tables make this method
challenging to implement.

An alternative---but apparently less well known---solution to this
problem was introduced by Moran at al.~\cite{moran2011visualization},
who observe that placing epsilon-sized boxes around all
\emph{vertices} of the input mesh, and ``snapping'' the corners of
those epsilon boxes to the centers of the cells they lie in (for which
they use a cell location kernel that walks the brick hierarchy), will
produce exactly the unstructured cells of the dual mesh.  This method
is more general, but apparently little known, arguably because it is
mentioned only as an aside in a paper with an otherwise different
focus.

\textbf{Contributions.} In this paper, we propose a method
that---though originally developed independently---can be viewed as a
simpler re-formulation of the Moran method. In particular, using a
well-defined integer labeling scheme for both cells and dual cells our
method avoid Moran's epsilon-offsetting, and reduces their cell
location kernel to simple binary search of integer indices; it is easy
 to implement, trivially parallelizable on a per-cell basis, and
lends itself naturally to a GPU implementation. Like Moran's method,
our method works on both octree AMR and structured AMR data (including
arbitrary level differences and ``holes'' in the model), and can be
used for either dual-mesh and/or iso-surface generation. We also
detail a CUDA reference implementation of this method (which will be
shared with this paper), and demonstrate that this can extract
crack-free iso-surfaces of even complex AMR data sets in a matter of
seconds (see Figure~\ref{fig:teaser}).

\section{Related Work}

Cell-centered AMR was first introduced by Berger and
Oliger~\cite{berger1984adaptive} and Berger and
Colella~\cite{berger1989local}, and in one or another form today is
used in a wide range of scientific simulation codes such as, for
example, Chombo~\cite{chombo}, Flash~\cite{flash}, LAVA~\cite{lava},
Rage~\cite{xrage}, and many others. In particular, though originally
introduced in the form of \emph{block-structured} AMR where
different-resolution grids are layered on top of each other, today
there are different variants ranging from block-structured to
octree-style AMR. Though the actual structure of the resulting AMR
mesh can vary wildly between those codes, they all follow certain
rules in that refinement happens in power-of-two factors, with data
values defined only for the center of each cell, and with certain
rules regarding how refined cells have to align to coarser
ones~\cite{berger1984adaptive,berger1989local}.

One of the first approaches to visualize AMR data was proposed by
Weber et al~\cite{weber2003extraction}, who proposed to first
construct the dual mesh, and extract an iso-surface from that the
Marching Cubes~\cite{lorensen87marching} case tables.  To construct
the dual messh Weber proposed to walk along the boundaries of the
input bricks, and ``stitching'' across level boundaries with elements
such as tetrahedra, pyramids, using a case table of some 36 different
cases to determine which element shape(s) to pick at any given
point. Notwithstanding some refinements
(e.g.,~\cite{weber2012efficient}) this method has remained the de
facto standard method for nearly two decades; its main downside is
that it requires some existing grid hierarchy, and in particular, that
models not contain any cases where more than two different levels
abut, as this would require significantly more complex case tables.

A significantly less known technique for computing this dual mesh was
proposed by Moran et al.~\cite{moran2011visualization}, in the context
of computing an interpolant for high-quality AMR volume rendering. In
their paper, Moran et al.~observe that there is a one-to-one
correspondence between vertices in the AMR mesh and unstructured cells
in the dual mesh on one hand, and input cell centers and dual mesh
vertices on the other---and that the dual mesh can thus be constructed
by surrounding each input mesh vertex with a tiny box whose vertices
then get snapped to the centers of the cells they lie in. To do this
snapping they proposed cell location kernel that walks a hierarchy of
AMR bricks. Though more general than Weber's, this technique seems
significantly less known, arguably because it was not primarily
proposed for iso-surface extraction. Though developed
independently---and with slightly different rationale---our method is
similar in spirit, and may actually best be viewed as a simpler
re-formulation of this technique.

A way of computing a indexed triangle mesh from a set of ``fat''
triangles has been previously proposed by
Bell~\cite{bell:10:indexedTriangles}, with a follow-up improvement by
Miller et al.~\cite{miller:topology}. The concept of dual contouring
has also been used, for example, by Nielsen~\cite{nielsen:DualMC},
Schaefer et al.~\cite{schaefer:dual-marching-cubes},
Kitware~\cite{kitware-blog}, and Carrard et
al~\cite{carrard:12:HyperTree}.

\begin{figure*}[ht]
  \resizebox{0.998\textwidth}{!}{
    (a)
    \includegraphics[height=3cm]{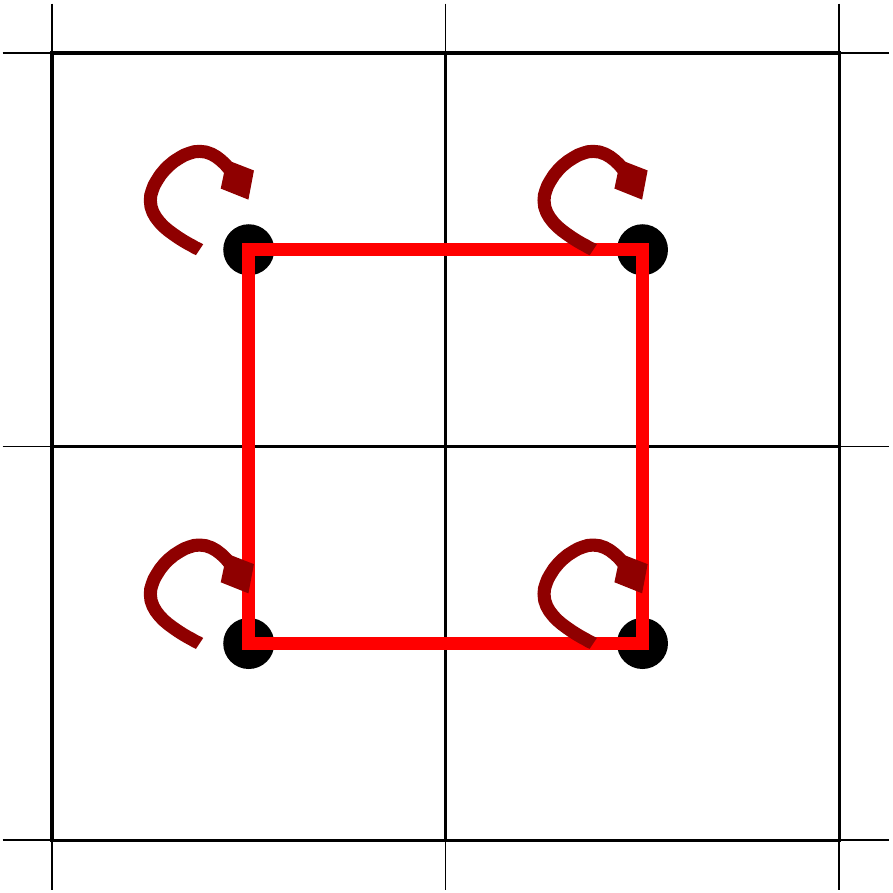}
    \quad
    (b)
    \includegraphics[height=3cm]{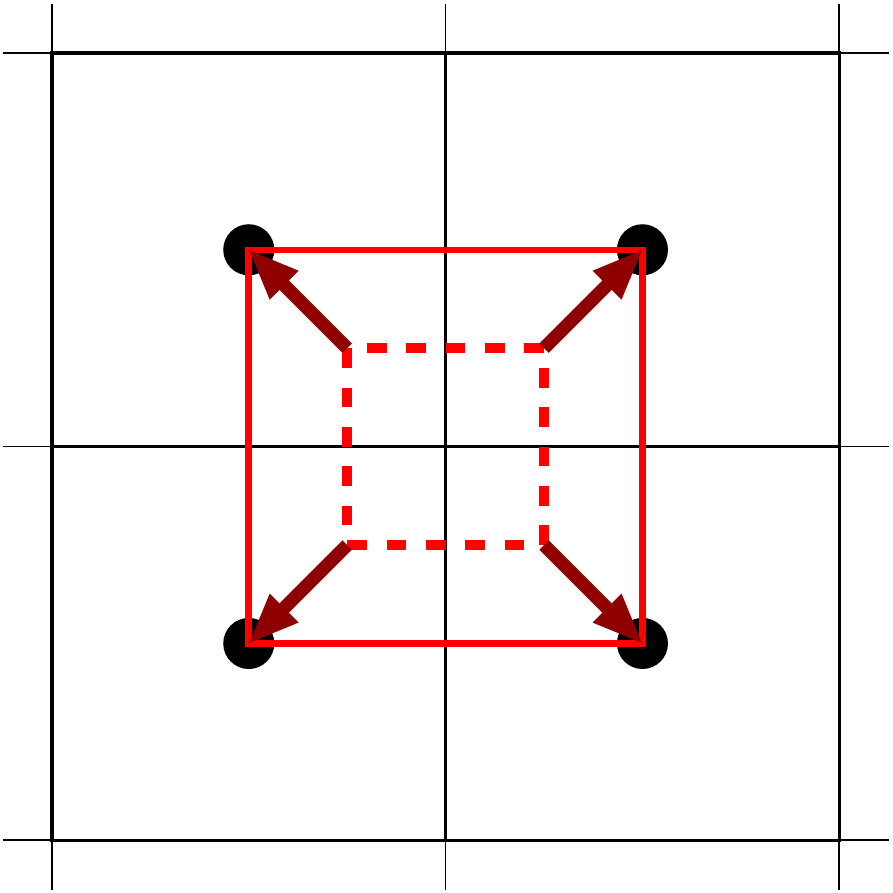}
    \quad
    (c)
    \includegraphics[height=3cm]{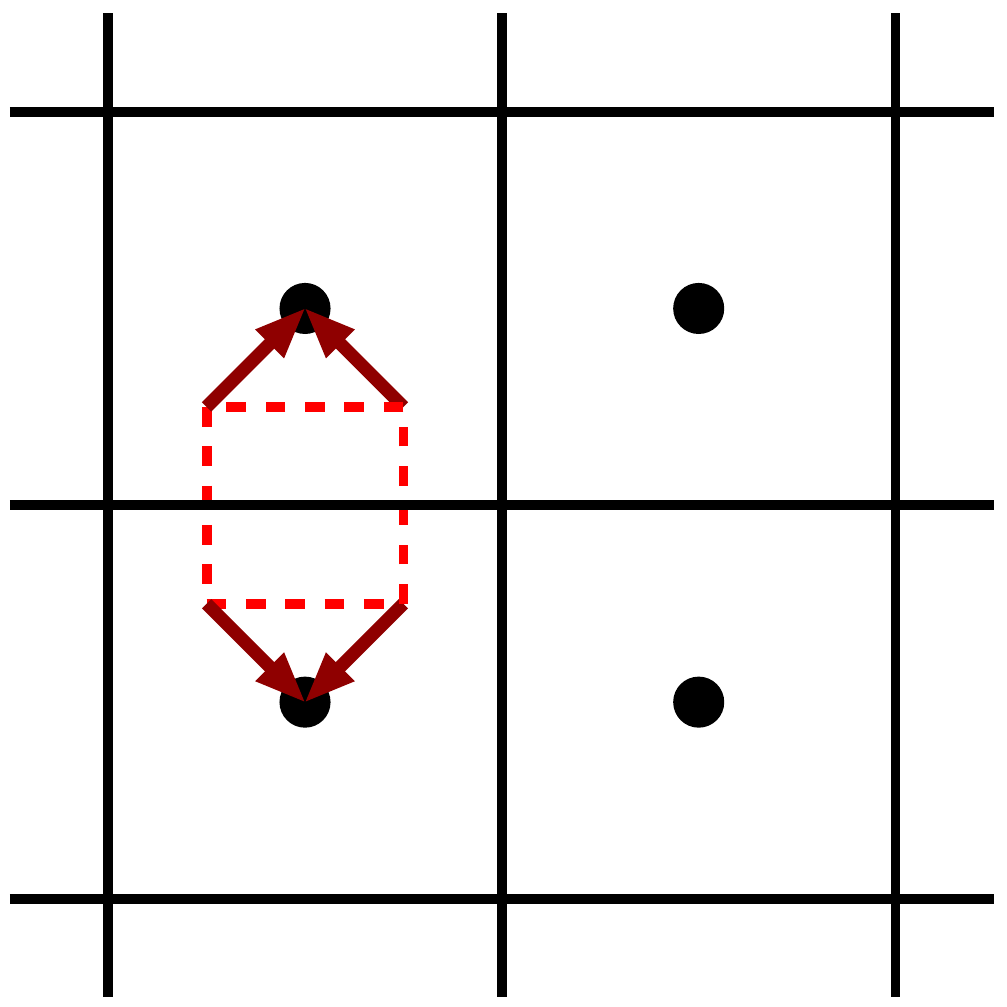}
    \quad
    (d)
    \includegraphics[height=3cm]{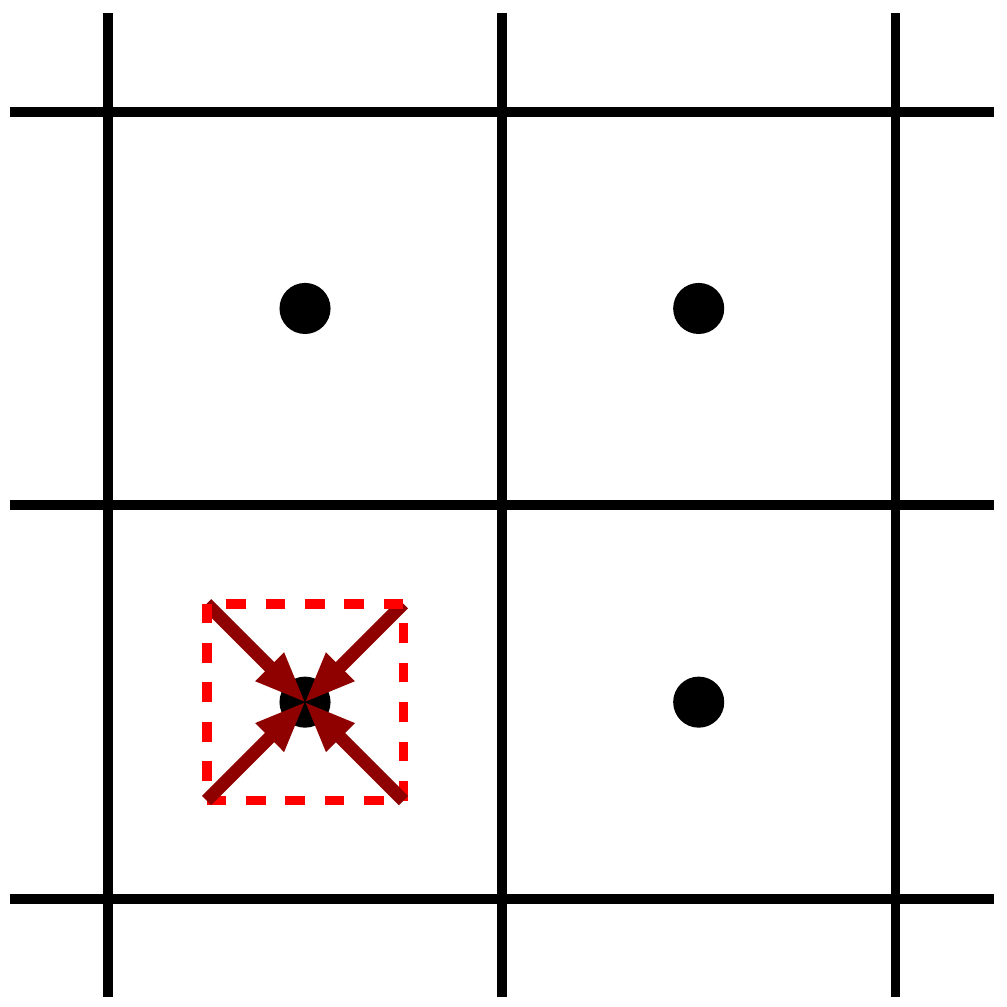}
    \quad
    (e)
    \includegraphics[height=3cm]{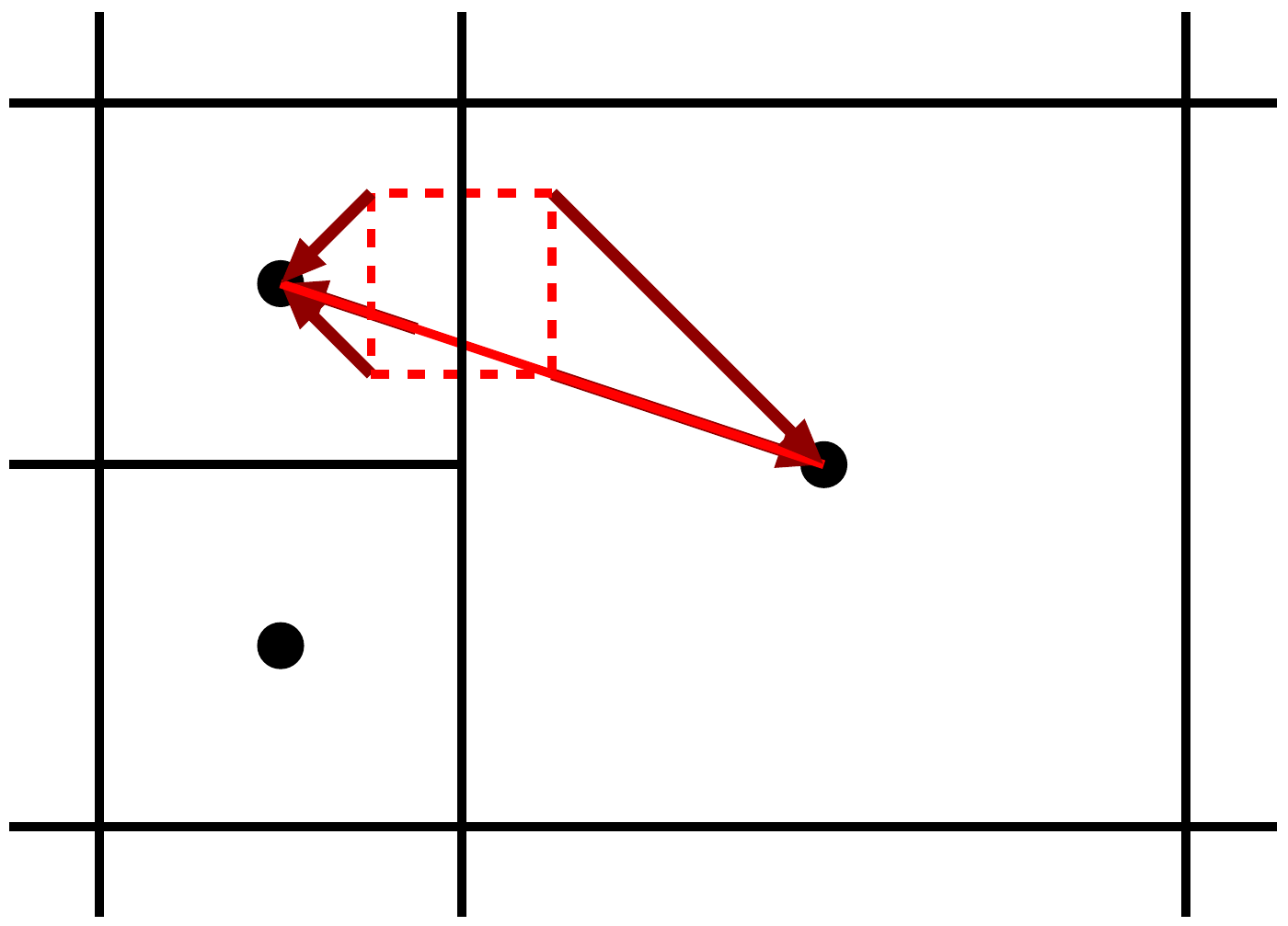}    
    \quad
    (f)
    \includegraphics[height=3cm]{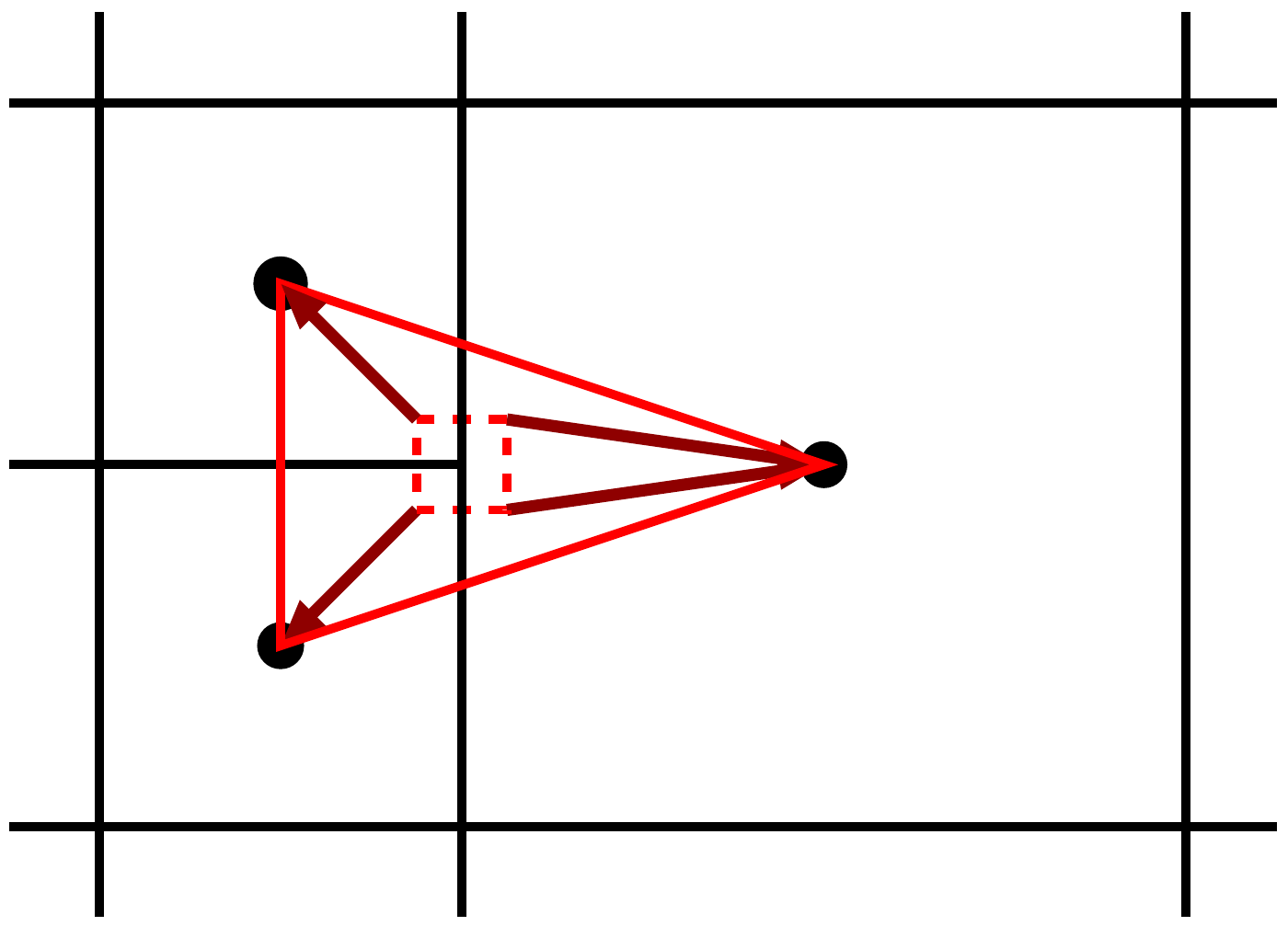}
    \quad
    (g)
    \includegraphics[height=3cm]{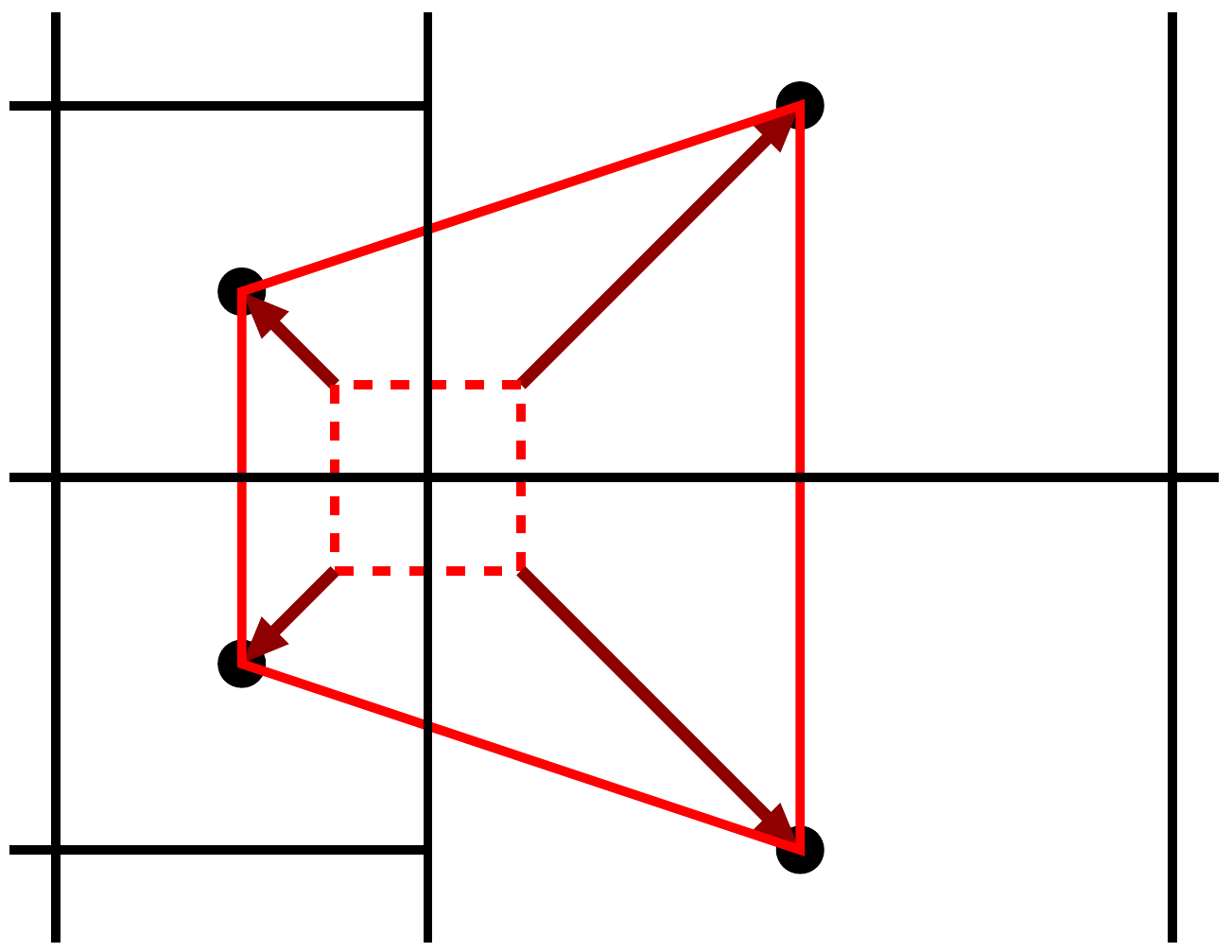}
  }
  \caption{\label{fig:cases} A 2D example of snapping various finest-level logical dual
    cells to actual cells (see text for a distinction between logical
    and actual cells). a) in a uniformly fine region, the vertices of
    fine duals snap to themselves, producing an actual fine dual. b-d)
    in a uniformly coarse regions, fine duals snap to a coarser real
    dual cell (b), degenerate lines (c), or degenerate points,
    depending on where they lie. e-g) fine(r) dual that straddle level
    boundaries snap to degenerate lines (e), triangles (f), or
    trapezoids (g). Ignoring all obviously degenerate cases (lines and
    points), snapping \emph{all} logical duals produces exactly the
    actual dual mesh.
    \vspace{-2em}
  }
\end{figure*}

\begin{figure}[ht]
  \resizebox{0.998\columnwidth}{!}{
    (a)
    \includegraphics[%angle=90,
      height=8cm]{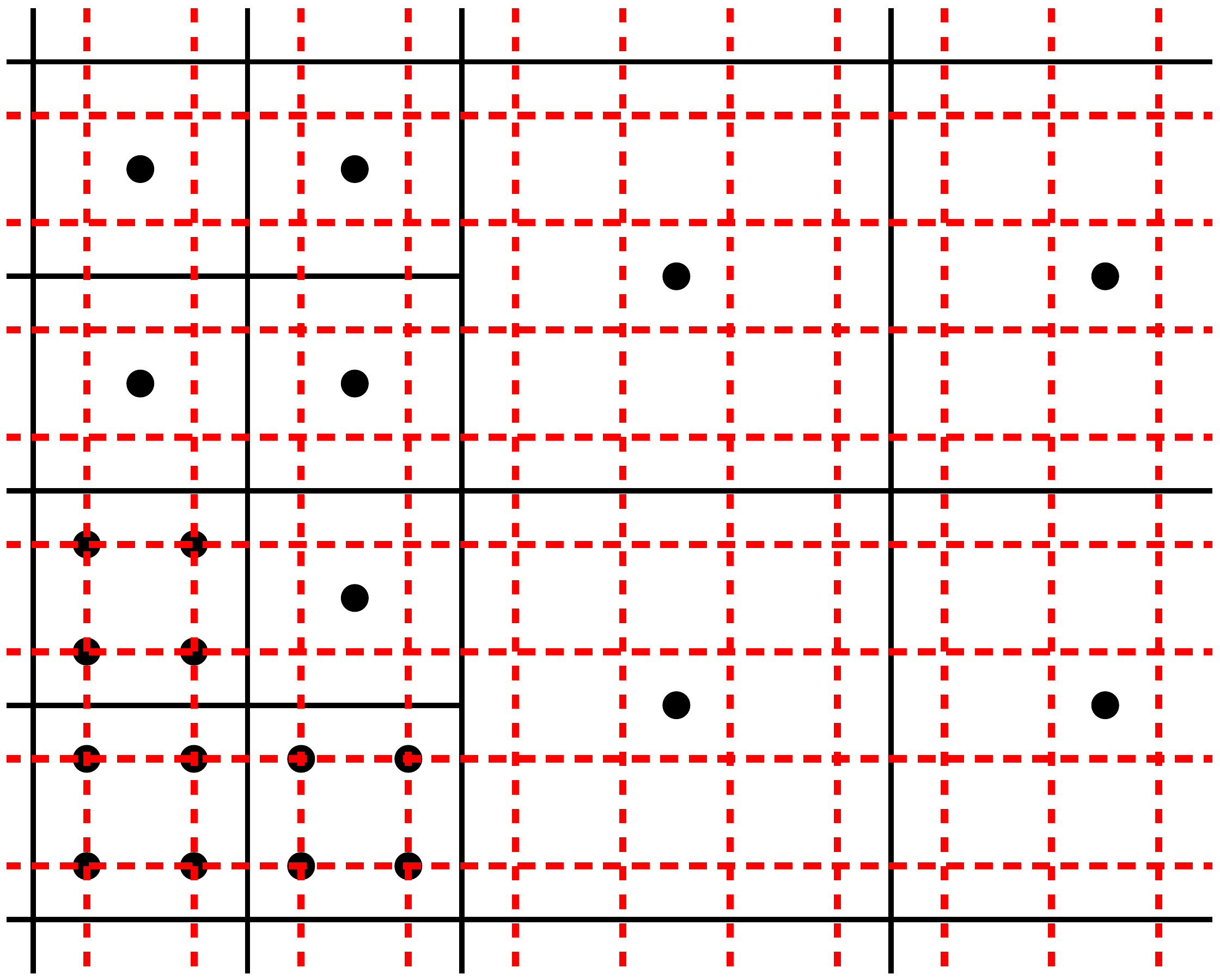}
    (b)
    \includegraphics[%angle=90,
      height=8cm]{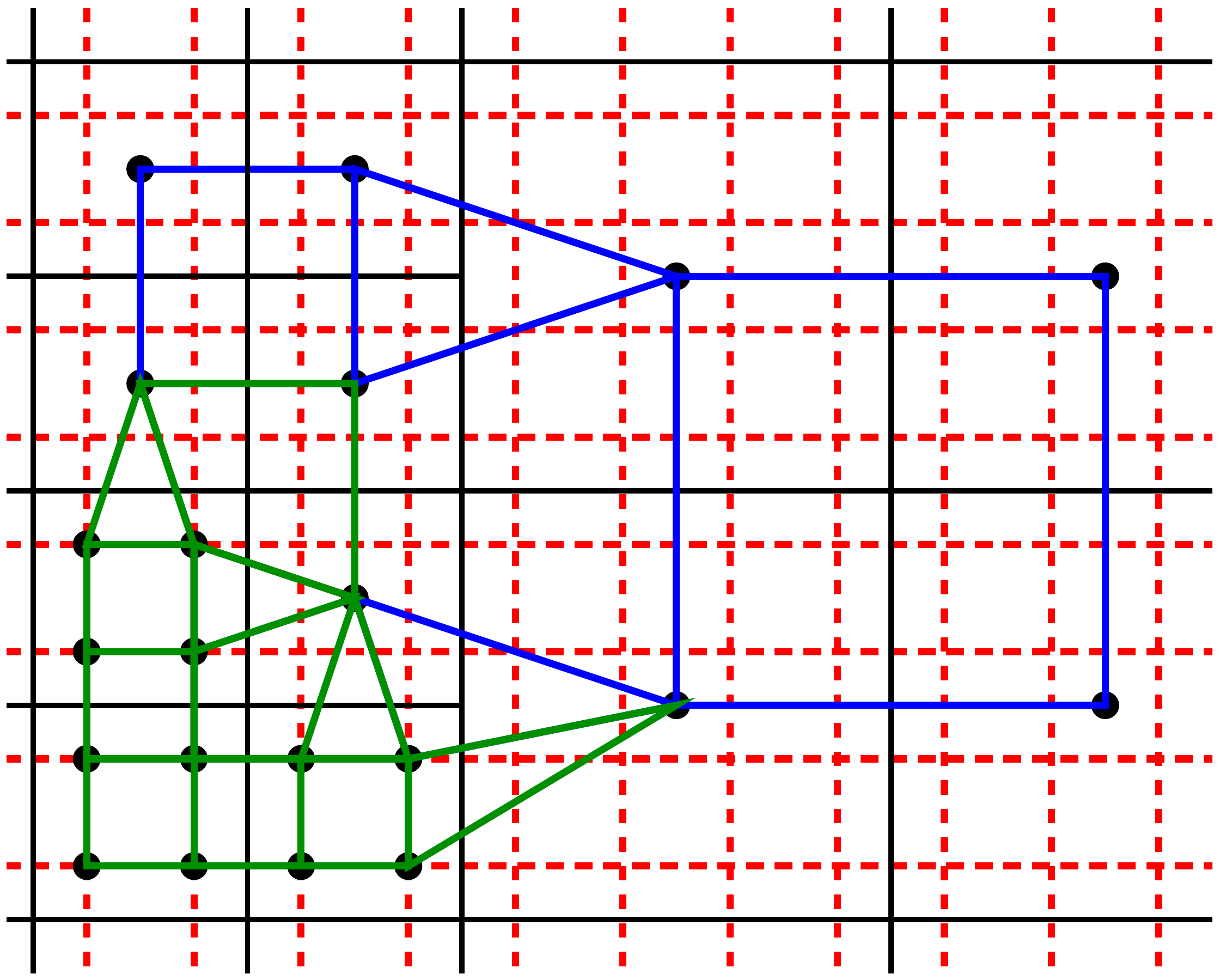}
  }\\[1ex]
  \resizebox{0.998\columnwidth}{!}{
    (c)
    \includegraphics[%angle=90,
      height=8cm]{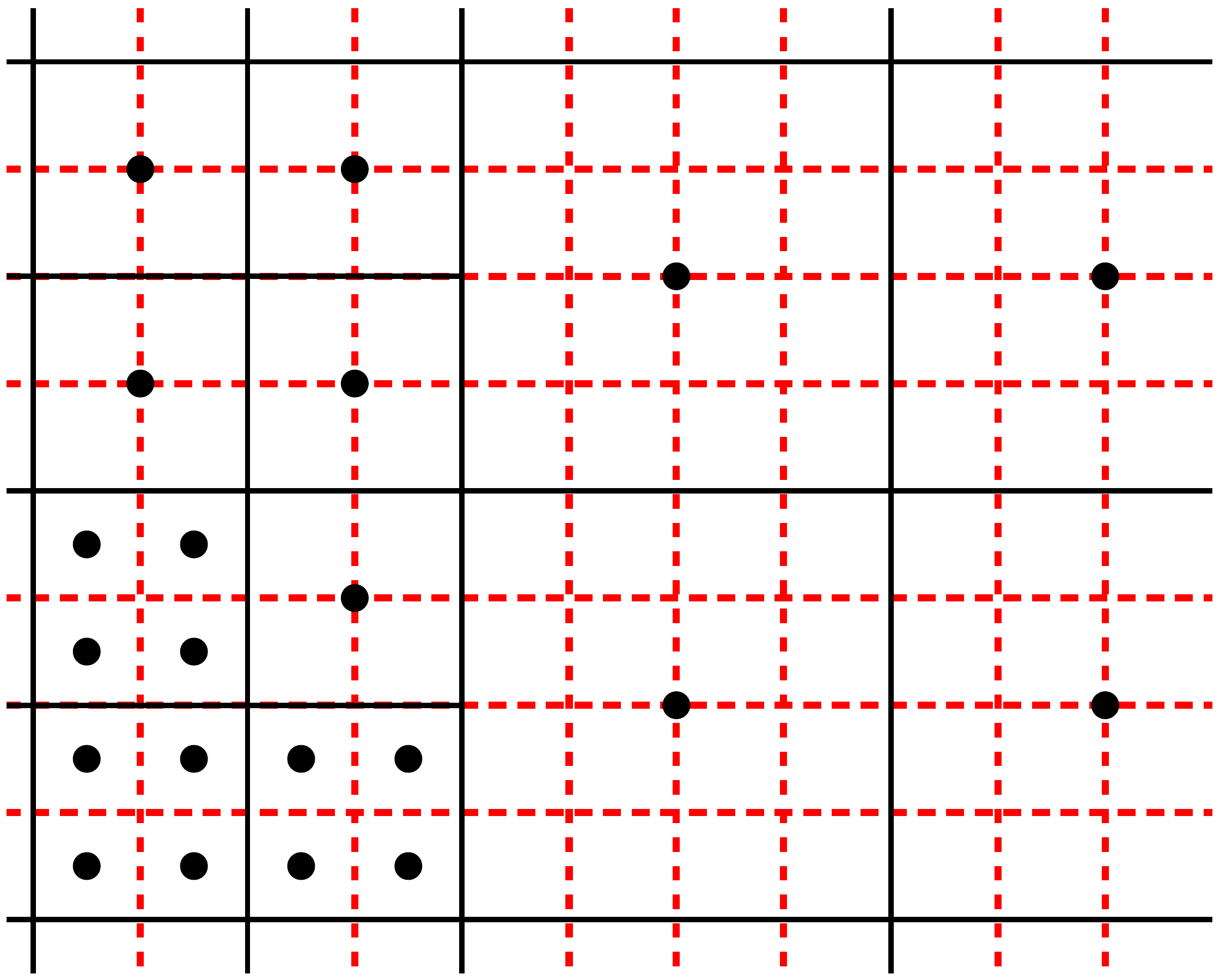}
    (d)
    \includegraphics[%angle=90,
      height=8cm]{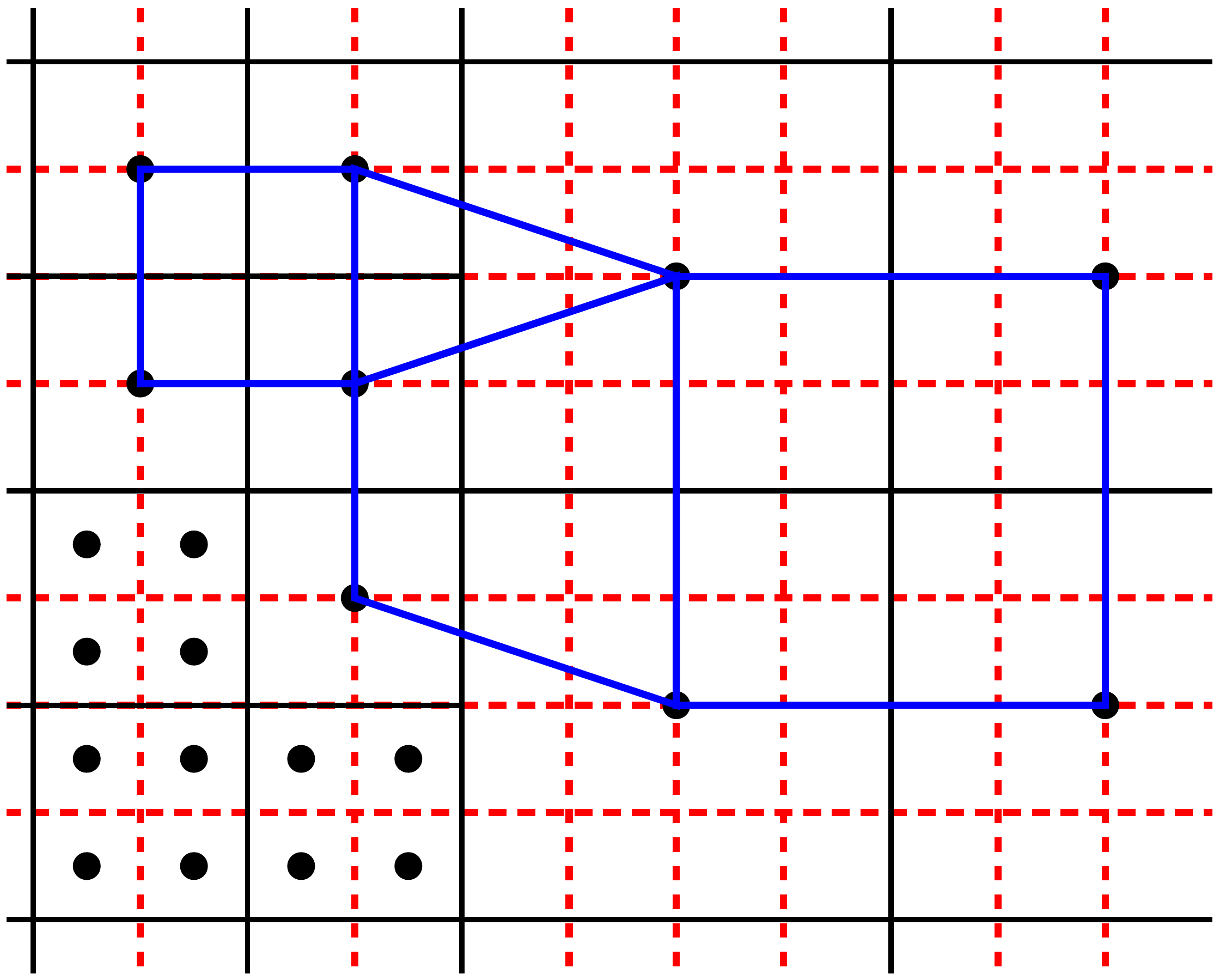}
  }
  \caption{\label{fig:different-dual-sizes}
    Illustration of snapping at different levels, on a sample
    mesh with 3 levels. a) the mesh with all finest-level dual cells
    (dotted red) b) the result of snapping all finest-level duals
    (which is the full dual mesh), with those shapes snap from at
    least one level-0 cell in green, and all others in blue.  c) The
    same mesh, with level-1 duals. d) The result of snapping these and
    discarding all shapes that snapped to a vertex \emph{finer} than
    level 1; which is clearly the same dual mesh for all regions level
    1 and coarser.
    \vspace{-2em} 
  }
\end{figure}

\section{Terminology and Prerequisites}

Though we use a different way of creating the dual mesh, at the root
out method works the same way as proposed by Weber: First construct
the dual mesh, then use that to extract the iso-surface. The key
contribution of this paper is the formulation of how the dual mesh is
constructed. To do this we first introduce a simple terminology for
logical AMR cells, dual cells, and levels, with which the final method
can be expressed in but a few lines of code.

\subsection{Cells, Dual Cells, and Terminology used in this Paper}

Throughout this paper we adopt a terminology where differentiate
between \emph{logical} and \emph{actual} AMR cells, where logical
cells refer to any cell that an AMR refinement count potentially
produce, and actual cells as those logical cells used by an actual AMR
data set. Due to the nature of AMR, all logical cells of any given
refinement level form an infinite structured grid, which we call a
``level''. In particular, throughout this paper we assume that level 0
is the finest level, with a cell size of 1; level 1 is the next
coarser level with cell size of 2, etc. I.e., cells on level $l$ have
a cell size of $2^l$, and have coordinates that are multiples of
$2^l$. To identify cells we refer to them through four coordinates
$(i,j,k;l)$, with ($i,j,k$) being that cell's lower-left corner, and
$l$ the level on which it lives. Ie, each logical cell
$C_{i,j,k}^{(l)}$ spans the space $(i,j,k)-(i+2^l,j+2^l,k+2^l)$ and is
centered at $\hat C_{i,j,k}^(l) =
(i+2^{(l-1}),j+2^{(l-1}),k+2^{(l-1}))$.

All logical level-$l$-cells form an infinite grid of $2^l$-sized
cells, which implied a similar logical dual grid of dual cells
$D_{i,j,k}^{(l)}$, which we define as spanning from $\hat
C_{i,j,k}^{(l)}$ to $\hat C_{i+1,j+1,k+1}^{(l)}$.

Using this terminology any given AMR data set $\mathfrak A$ can then
be descibed by which \emph{actual} cells it contains, and what data
value(s) these cells carry. Note in particular that we do not assume
any existing block structure or hierarchy; just that we know which
cells $(i,j,k;l)$ a given data set contains.

\subsection{Efficient Cell Location}
\label{sec:cell-location}

Given such a data set, in the later stages of our algorithm we assume
that there will be a fast \emph{cell location} operator
\textsc{snap(C)} that finds, for any given set of cell coordinates
$(i,j,k;l)$, the \emph{actual} input cell $\hat C_{i',j',k'}^{(l')}$ that
contains the coordinates $(i,j,k)$, or, if no such actual data set
cell exists, a special ``does not exist'' indicator $\emptyset$. Note
in particular that the actual cell for the queried coordinates
$(i,j,k;l)$ does \emph{not} have to have these logical cell
coordinates, but can absolutely be on either a finer or coarser level.

In our implementation we implement this operator by first sorting the
set of input cells by their integer coordinates $(i,j,k)$, and by
determining the maximum number of levels $L$. With this sorted array
we can then quickly check if the data set contains a given cell with
coordinates $C_{i,j,k}^{(l)}$ by simply running a binary search, for
which in our implementation we use \emph{thrust}'s
\code{thrust::lower\_bounds}~\cite{thrust}.

For any given integer position $(i,j,k)$ we can then find the actual
cell containing these coordinates---if it exists---by simply iterating
over all levels; for each level $l$ we first project $(i,j,k)$ to
valid level-$l$ cell coordinates by zeroing the lower $l$ bits of $i$,
$j$, and $k$, then perform the binary search. If any tested level
finds an actual cell with these coordinates this is the cell we have
been looking for; otherwise we return a $\emptyset$. Of course, if we
have a good guess on what level the queried cell might be on we can
test this level first, saving some additional searches. Note in
particular that we never need any floating-point coordinates,
epsilons, or hierarchies, only pre-sorted integer cell
coordinates. Furthermore, we make no distinctions between ``inner''
and ``boundary'' cells whatsoever; all cells are treated exactly the
same, simplifying the algorithms and aiding parallelism.

\section{Computing the Dual Mesh using Degenerate Dual Cells}
\label{sec:dual-generation}

Using this terminology and ``cell location'' kernel we now make the
same argument as already made by Moran et al, namely, that
\emph{snapping} the vertices of logical dual cells to the centers of
the actual cells they lie in will produce exactly the unstructured
dual cells.

%% The core approach of this paper is motivated by three
%% observations. First, that with our cell location operator we can
%% \emph{snap} any point $P$ to an actual vertex of the dual mesh by
%% finding the actual cell $\mathcal A(P)$ containing $P$, and moving
%% it to the center of that cell. Second, that we can similarly take
%% any 3D oblong $H$ and snap it to the dual mesh by snapping its 8
%% vertices, generating a \emph{potentially degenerate} hexahedron
%% $H'$; and third, that when projecting the right input hexahedra
%% their projections will form exactly the dual mesh. The

\subsection{Conceptual Construction on the Finest Level}

To illustrate the latter, let us first consider the 2D case,
and at least for now, let us consider all(!) finest-level dual
cells. In Figure~\ref{fig:cases} we have illustrated several cases on
how an originally square-shaped finest-level dual snaps in different
regions of a data set: In a finest-level region that square just snaps
to itself (a); in coarser regions it either snaps to a larger square
(b), to a line (c), or a point (d), depending on whether the dual
square covers a coarser cell vertex (b), straddles and edge (c), or
lies within a coarser cell (d). 
%In particular, we point out that the
%vertices of a level-$l$l dual cell can never directly \emph{on}
%vertices, edges, or faces of cells that are level $l$ or coarser.

Along level boundaries, the shapes produced by snapping get more
diverse, but still follow the same concept: in 2D, snapping a
finest-level dual that straddles a level boundary can also produce a
triangle (e) or trapezoid (f) along a simple boundary between only two
levels; and a general quadrilateral if more than two different levels
are involved (g). In 3D exactly the same construction applies, except
that the shapes produced get even more diverse by adding another
dimension: generally speaking, in the simple case inside a homogeneous
regions 2D squares becomes 3D cubes, but along boundaries triangles
and trapezoids can become tetrahedra, pyramids, wedges, or general
hexahedra. Some of these shapes are non-trivial, as some of the
hexahedra, wedges and pyramids can have faces that are non-flat
bilinear patches.

In particular, we point out that this snapping does \emph{not} make
any assumptions as to how many different levels are involved, how big
the level differences are, etc. Obviously, if any of the vertices does
not snap to any actual cell the shape is not part of the dual mesh,
and can be ignored.

\subsection{Efficient Construction}

Though constructing the dual mesh by snapping all finest-level cells
does indeed work, for complex models it would be very inefficient to
do so: for example, for the NASA Landing Gear Model outlined in
Figure~\ref{fig:models}, the logical finest-level grid contains
$131K\times 131K\times 65K$ cells, translating to roughly $1.1$
quadrillion finest-level duals that we would need to snap.

However, as already pointed out by Moran et al.~all that is required
to find all shapes is that the snapped shapes cover all vertices of
the input mesh. We further observe that every vertex of the input mesh
is by definition one of the eight vertices of an actual cell
$C_{i,j,k}^{(l)}$, and as such, at the center of one of the eight
logical dual cells $D_{i+\Delta i,j+\Delta j,k+\Delta k}^{(l)}$
($\Delta i, \Delta j, \Delta k \in \{0,1\}$) that cover this cells.
This suggests that all we have to do is to generate, for each actual
input cell $C_{i,j,k}^{(l)}$, the coordinates of its 8 dual cells, and
snap those.

\subsection{Avoiding Duplicates}

This method of snapping every actual cell's eight duals is simple and
correct, but since dual cells are generally shared among many
different actual cells would naively generate each output cell several
times. One way to avoid this would be to explicitly tag vertices that
have already been processed, but this would require additional
bookkeeping that would complicate the algorithm and in particular
hamper parallelism.

Fortunately, we can easily avoid duplicates using a set of three
simple rules that disambiguate which logical dual cell $D$ is allowed
to produce the shape:
\begin{description}
\item[rule \#1] if any of the corners of the dual does not snap to any
  actual cells, then this dual does not produce a valid shape at all
\item[rule \#2] if \emph{any} of the corners of dual cell snaps to a
  \emph{finer} level, then that same shape will already be produced on
  that finer level, and will not be emitted.
\item[rule \#3] if any of the eight vertices of $D_{i,j,k}^{(l)}$ snaps
  to a cell $(i',j',k')'$ that is on the same level $l$, but has with
  coordinates lower than $(i,j,k)$, then that same shape will already
  be produced by dual $D_{i',j',k'}^{(l)}$, and will not be emitted.
\end{description}
These conditions are trivial to test with just a few integer
compares, and guarantee that every dual shape will get emitted exactly
once.

\subsection{Pseudo-Code}

Though the previous sections was intentionally verbose, the resulting
algorithm is intriguingly simple. For each of a cell's 8 dual cells we
construct the eight vertices, snap them, and reject the generated
shape if it violates any of the previous three simple rules:
%is either at a boundary to the outside, at a finer-level
%boundary, or already generated by another cell:%

{\relsize{-1}{
%% \begin{lstlisting}
%% void doCell(i,j,k,l) {
%%   Cell self(i,j,k,l);
%%   for (int dk in (-1,0) ) {
%%     for (int dj  in (-1,0) ) {
%%       for (int di in (-1,0) ) {
%%         doDualCell(self,i+di,j+dj,k+dk,l)
%% }}}}
\begin{lstlisting}
void doCell(i,j,k,l) {
  Cell self(i,j,k,l);
  for (int di,dj,dk in (0,1) ) 
     doDualCell(self,i+di,j+dj,k+dk,l)

void doDualCell(self,i,j,k,l) { 
  // the eight vertices we're snapping to:
  Cell vertex[2][2][2];
  for (int di,dj,dk in (0,1) ) 
    Cell v = snap(i+di,j+dj,k+dk);
    if (v == invalidCell) 
      // dual reached outside the mesh -> reject
      return;
    if (v.level < l) 
      // dual cell straddles to a finer level,
      // finer cell will generate this -> reject
      return
    if (v.level == l && v < self) 
      // 'v' will generate the same dual,
      // and is smaller that we are -> reject
      return
    // store this vertex:
    vertex[dk][dj][di]
  // now *have* a (potentially degenerate) dual cell:
  emitDualCell(vertex[][][]);
\end{lstlisting}
}}

We observe that the entire construction is completely parallel,
without any locks or synchronization whatsoever. In our actual CUDA
implementation (see below) we launch one thread per each of the 8 dual
cells for any of the inputs, and simply run the above code, using the
fast cell location kernel outlined above.

\section{Application to Iso-Surface Extraction}
\label{sec:iso-extraction}

The construction discussed so far is general, and generates the dual
mesh irrespective of its application to iso-surface extraction. As
done by Moran et al, the resulting dual mesh, could, for example, be
stored, and used to create an interpolant for direct volume rendering.
However, an even easier application is iso-surface extraction. For
this, the key observation is that though some of the snapped dual
cells may degenerate to wedges, tetrahedra, pyramids, or even just
lines or points we can still treat \emph{all} of those shapes (even
the degenerate ones) as if they were hexahedra, and simply feed those
into the marching cubes (MC) algorithm.

That some of those hexahedras' vertices have snapped to the same cell
does not matter for the MC case tables at all: MC generates all of its
surface vertices along edges of the hexahedra that have one vertex
above and one below the desired iso-surface; in the case where edges
snap to a single vertex both endpoints of that ``edge'' will by
definition be the same, so triangles will never be generated along
collapsed edges. Similarly, even if a face snaps to a non-trivial
shape such as a triangle or even a bi-linear patch we know that both
duals that share this face will have generated it in exactly the same
way, thus the MC tables will place the respective shape's vertices
along that face at exactly the same locations, thus guaranteeing that
no cracks can occur.

We observe that once again the entire process in our formulation is
intriguingly simple: start one thread per each of the duals of the $N$
actual input cells, have that thread compute ``its'' dual cell, and if
found, feed the resulting eight vertices---denegerate or not---into
the standard MC case tables, then simply check which of the up to
three generated triangles have three different vertices. As evidenced
by the accompanying CUDA code the entire algorithm can be formulated
in about as many lines of code as taken up by the MC case tables.

\subsection{CUDA Implementation}

\begin{figure*}[h]
  \centering
  \setlength\tabcolsep{0.1ex}
  \relsize{-1}{
  \begin{tabular}{cccc}
    \includegraphics[width=.24\textwidth]{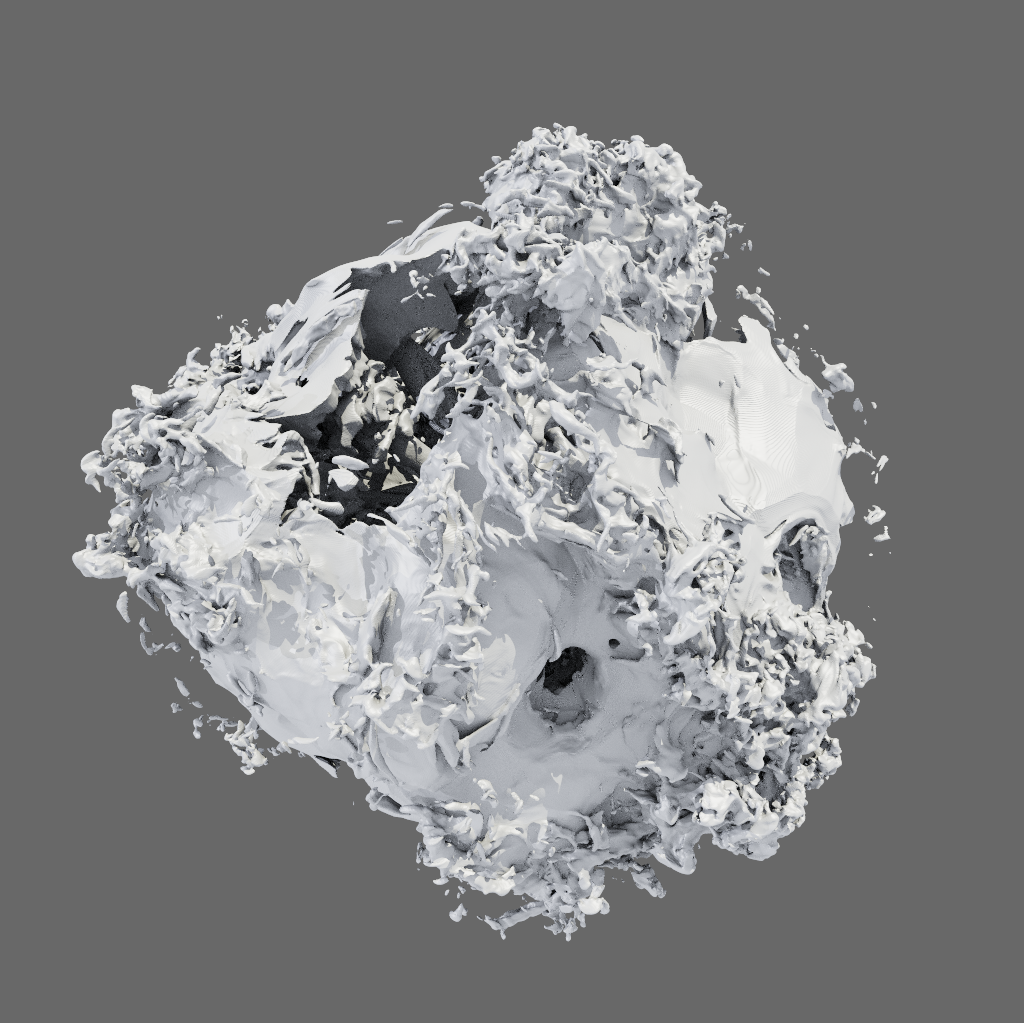}
    &    
    \includegraphics[width=.24\textwidth]{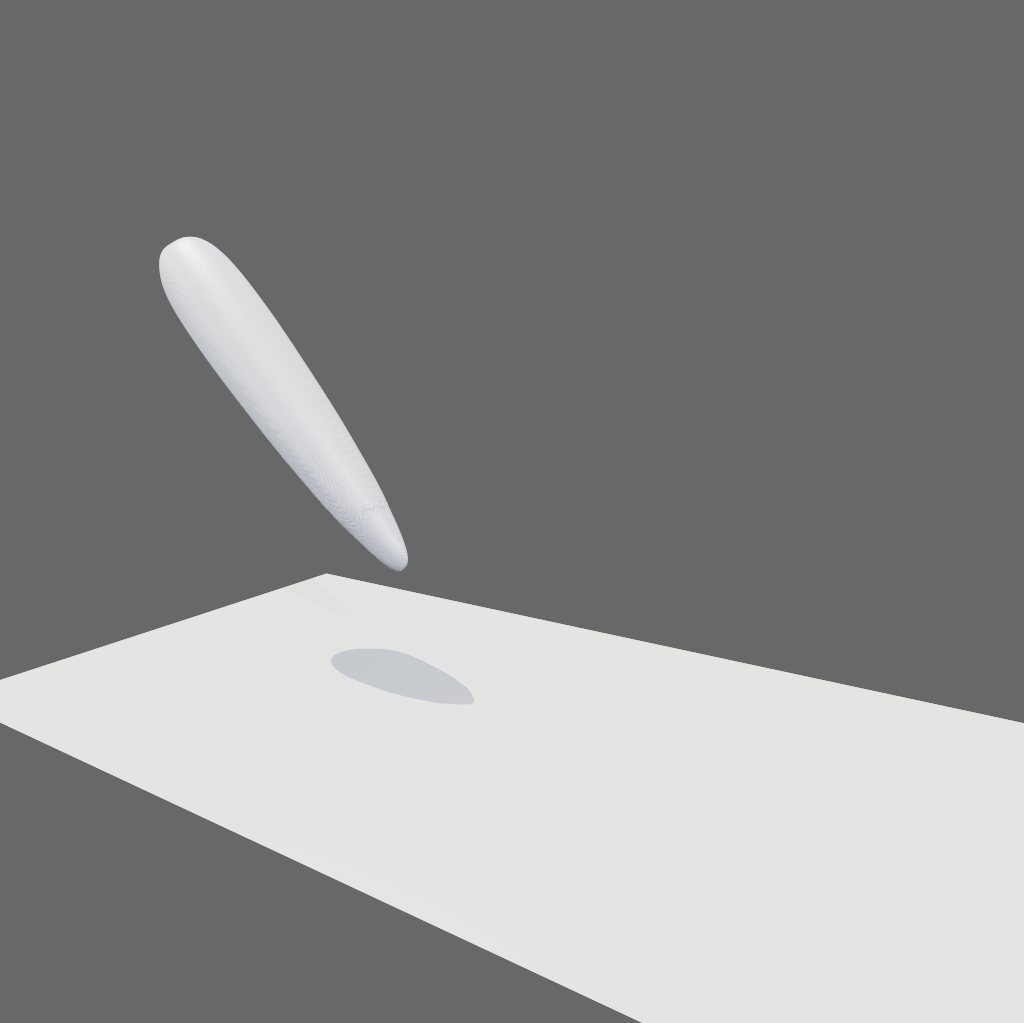}
    &    
    \includegraphics[width=.24\textwidth]{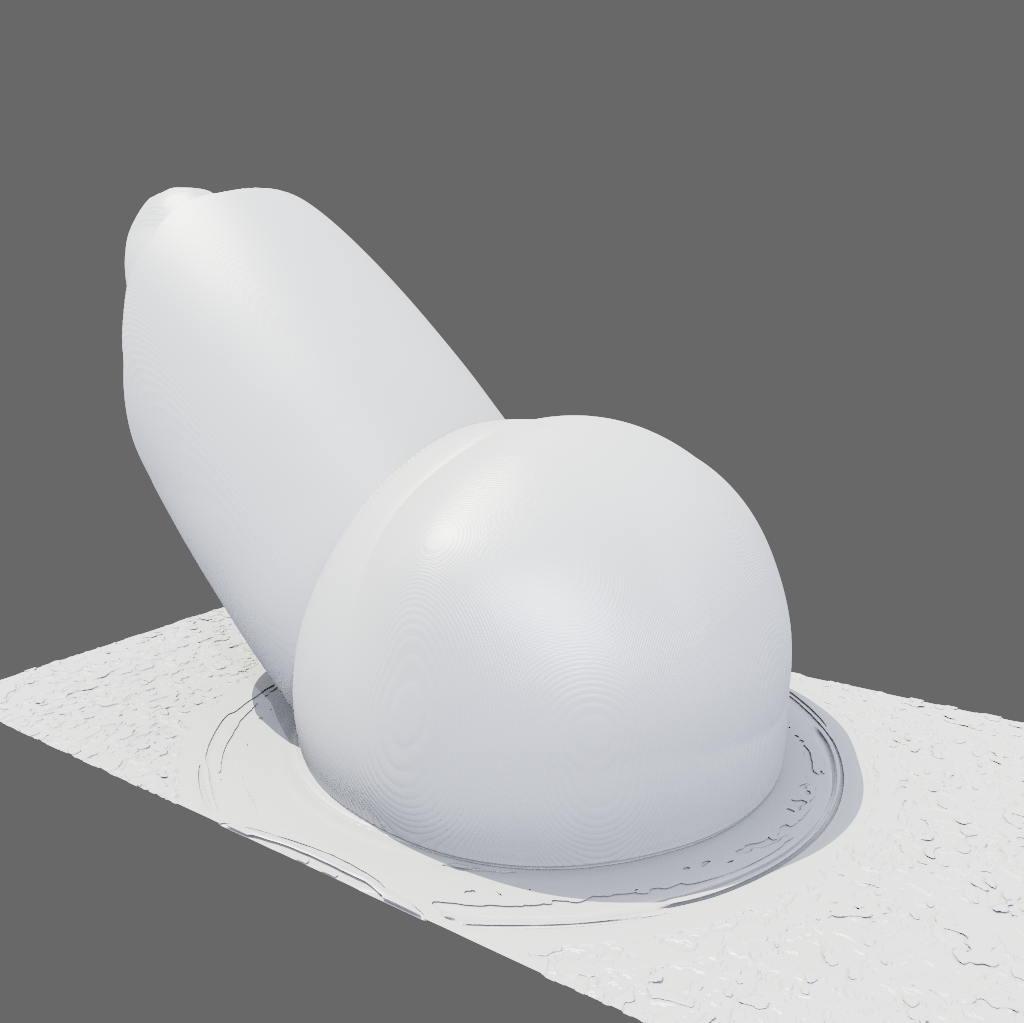}
    &    
    \includegraphics[width=.24\textwidth]{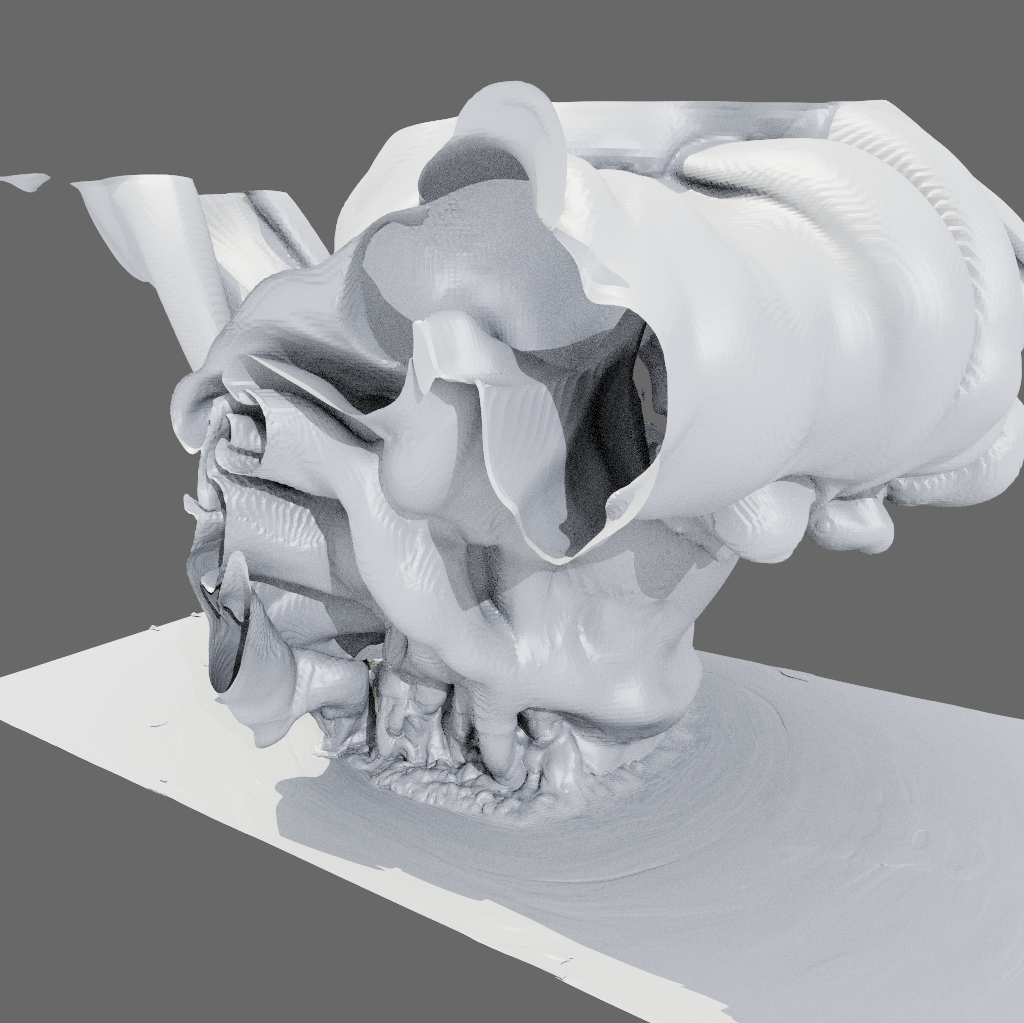}
    \\
    \emph{stellar cluster wind}
    &
    \emph{impact-5700}
    &
    \emph{impact-20060}
    &
    \emph{impact-46112}
    \\
    (\emph{flash}, block-structured)
    &
    (\emph{xRage}, octree-AMR)
    &
    (\emph{xRage}, octree-AMR)
    &
    (\emph{xRage}, octree-AMR)
    \\
    4 levels, 77M cells
    &
    4 levels, 26M cells
    &
    4 levels, 151M cells
    &
    4 levels, 270M cells
    \\
    $||v||$ field, $\rho=9.5M$
    &
    $tev$ field, $\rho=0.22$
    &
    $tev$ field, $\rho=0.22$
    &
    $tev$ field, $\rho=0.025$
    \\
    24.3M triangles, 5.7/14.1 sec
    &
    308K triangles, 0.38/0.91 sec
    &
    2.3M triangles, 2.3/5.5 sec
    &
    5.1M triangles, 3.8/9.4 sec
    \\
    \includegraphics[width=.24\textwidth]{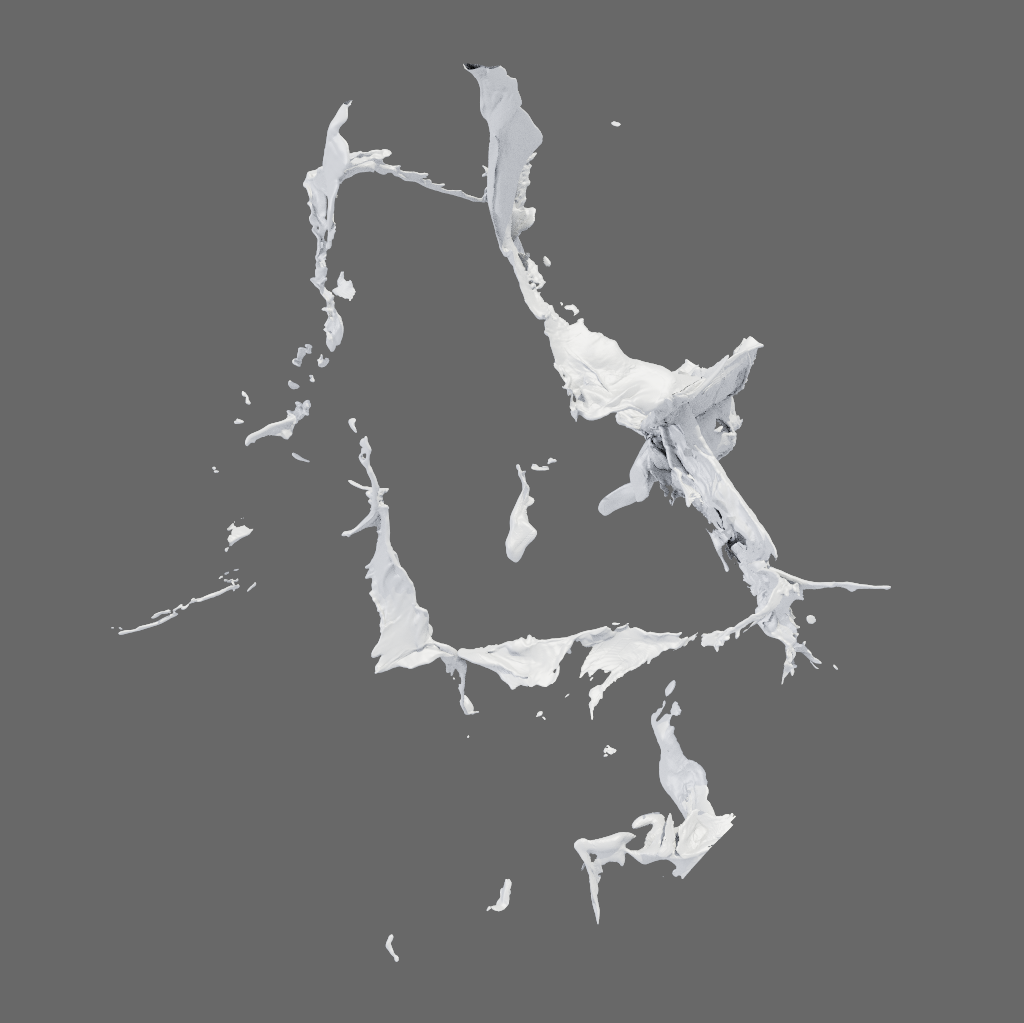}
    &    
    \includegraphics[width=.24\textwidth]{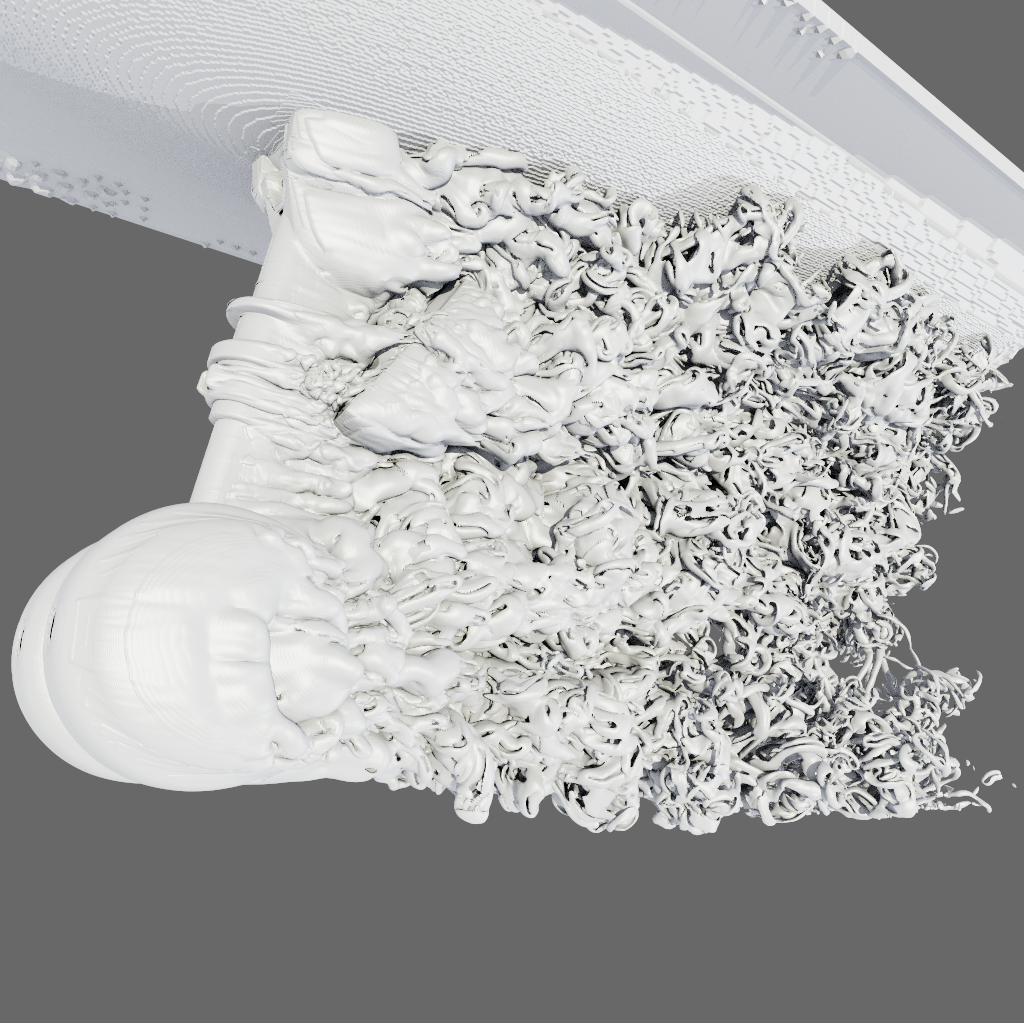}
    &    
    \includegraphics[width=.24\textwidth]{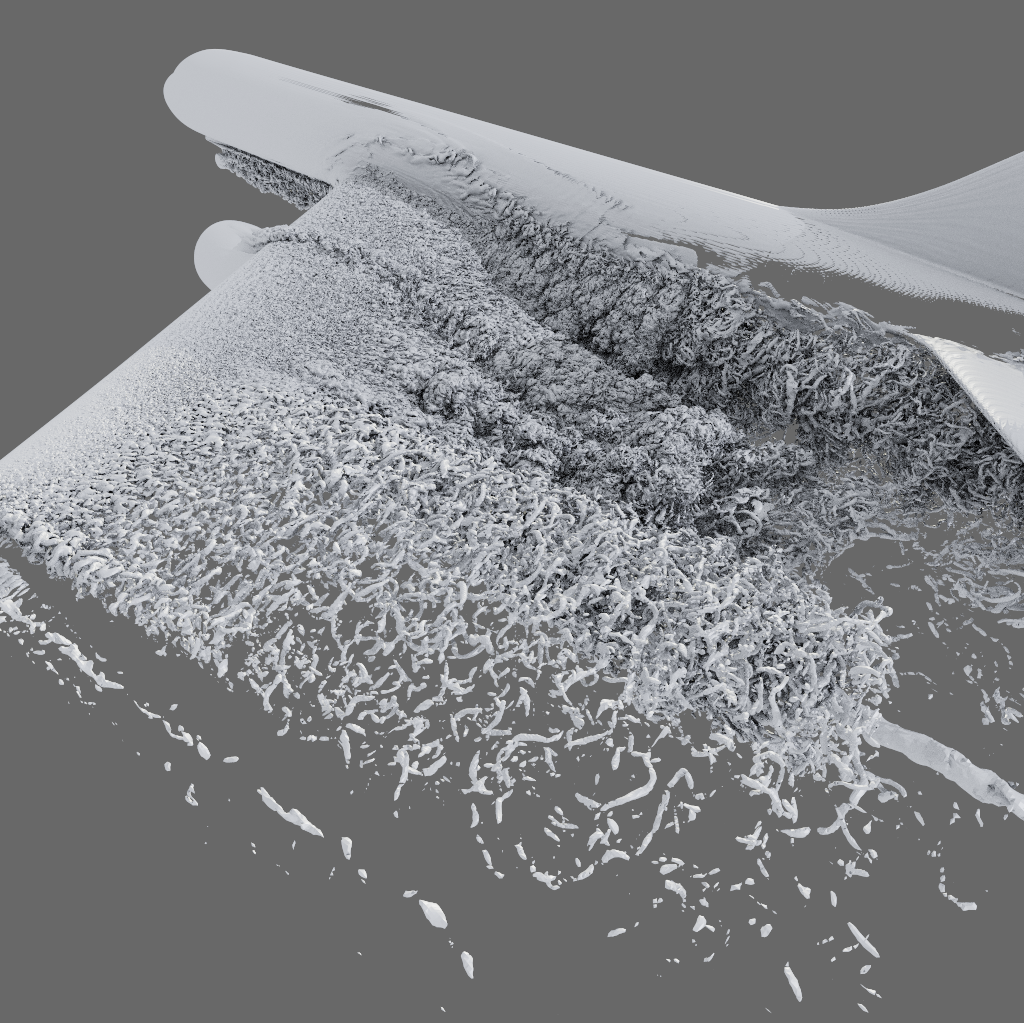}
    &    
    \includegraphics[width=.24\textwidth]{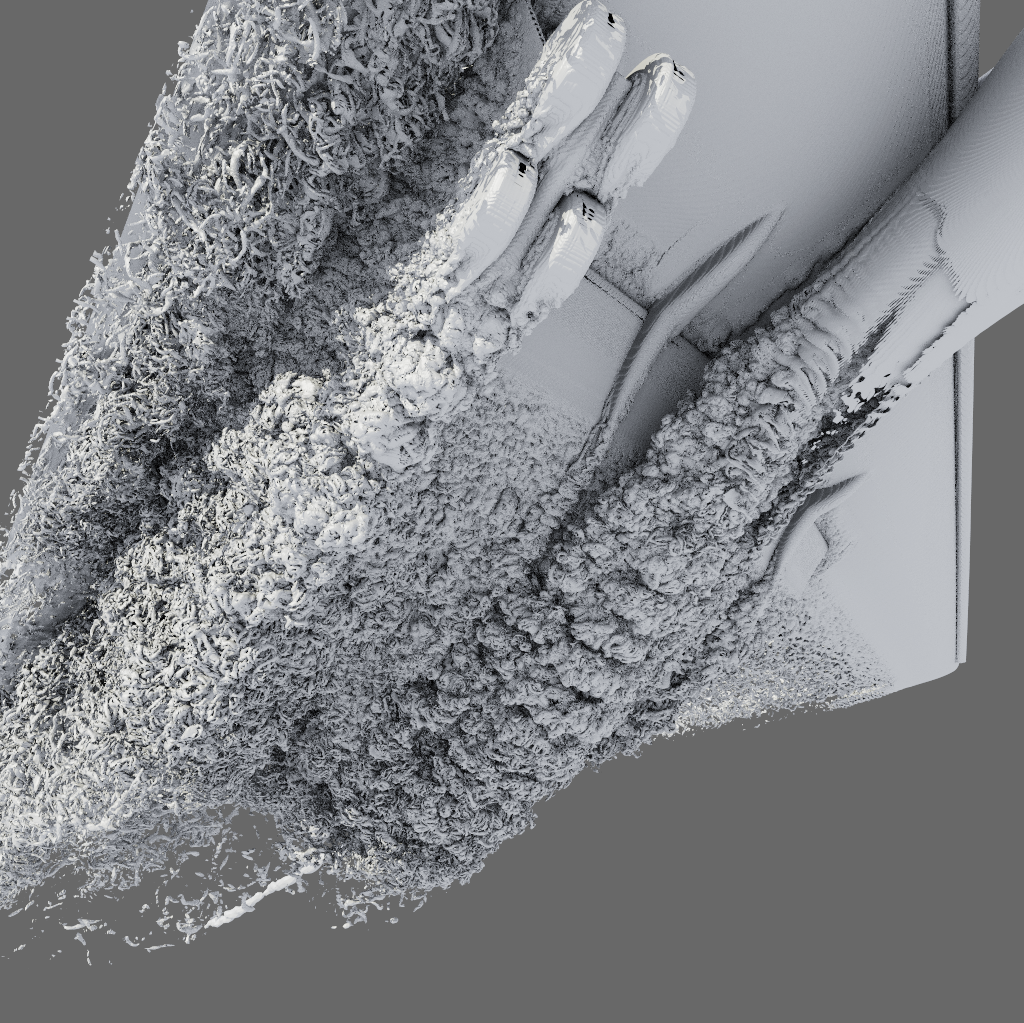}
    \\
    \emph{TAC molecular cloud}
    &
    \emph{NASA landing gear}
    &
    \multicolumn{2}{c}{\emph{NASA exajet}}
    \\
    (\emph{flash}, block structured)
    &
    (\emph{LAVA}, block structured)
    &
    \multicolumn{2}{c}{Exa PowerFlow, Octree-AMR}
    \\
    5 levels, 392M cells
    &
    13 levels, 249M cells
    &
    \multicolumn{2}{c}{4 levels, 626M cells}
    \\
    $temp$ field, $\rho=20$
    &
    $||vorticity||$ field, $\rho=4000$
    &
    \multicolumn{2}{c}{$||vorticity||$ field, $\rho=1000$}
    \\
    4M triangles, 1.1/2.6 sec
    &
    52.7M triangles, 3.4/8.9 sec
    &
    \multicolumn{2}{c}{130M triangles, 8.4/22.2 sec}
    \\
  \end{tabular}
  }
  %\vspace{-.8em}
  \caption{\label{fig:models} The data sets we used to evaluate
    and validate our method, each with the type of input AMR (block structured vs octree AMR), the code that generated it, number of input cells, generated number
    of triangles, and timings using our method.
%    : a) a simulation of \emph{stellar
%    cluster wind} from Princeton, in \emph{flash} block-structured AMR, b-d)
%    three time steps of the LANL
%    Asteroid Impact simulation~\cite{impact-dataset} (\iw{TODO}),
%    e) the \emph{TAC Molecular Cloud}
%    data set from the Theoretical Astrophysics Group 
%    Cologne (\emph{chombo} block-structured,
%    f)~the \emph{NASA Landing Gear} also used by Wang et
%    al~\cite{wang:18:iso-amr} (LAVA, block-structured), and
%    g+h) the \emph{Exajet} (Octree-AMR). For each
%    data set, we specify the number of AMR cells in the data set,
%    as well as the number of triangles generated for the
    %    given field and iso-value.
    Timing data refers to our CUDA implementation on
    a RTX~8000 GPU; timings given as X/Y seconds means that it
    took X seconds for the core surface generation kernel,
    and Y seconds total for an entire pipeline including data
    upload to the GPU, sorting for cell location, two passes for first
    counting and then generating the triangles, converting the
    triangles to an indexed face set, and downloading back to the
    host.  }
  \vspace{-2em}
\end{figure*}

To demonstrate this application we implemented both our dual-cell
generation and iso-surface extraction in CUDA. We split out code into
the cell location kernel outlines above
(Section~\ref{sec:cell-location}), a kernel that executes the dual
cell generation using this snapping kernel
(Section~\ref{sec:dual-generation}), and a kernel that implements the
MC case table for those duals that passed all tests. For the MC step,
we used the case tables and edge tables from VTK~\cite{VTK}, and put
those into CUDA \textsc{\_constant\_} memory for fast access.

The kernel then takes the 8 vertices it is fed with, computes the 3D
position at the center of each cell, and looks up the respective
cell's scalar values. Just like in the regular MC case the resulting 8
scalars are used to construct the case table index, which tells us
which triangle vertices to construct on which edges of the hexahedron,
and which triangles to emit. Obviously, when MC tags an edge for
creating a vertex, we interpolate this vertex from the snapped
positions, but otherwise there is no difference whatsoever; in fact,
the code is an almost literal copy-and-paste from VTK's
implementation.

\subsection{Generation of Output Triangle Mesh}

Every time our CUDA kernel runs the MC step, it can produce up to 5
triangles. We use an atomic counter to append each such triangle to a
buffer provided to our kernel, each time checking if the buffer is
actually large enough. As the algorithm is completely deterministic,
this means we can use exactly the same code both for counting how many
triangles will eventually be generated, and for actually generating
them: In a first pass, we execute the algorithm with a zero-sized
outside buffer, which will not write any triangles, but modify the
atomic to tell us how many \emph{would} have been written. We then
allocate an output buffer of that given size, reset the atomic, and
simply run the same code again.

The result of this pass is a list of a ``fat'' triangle mesh where
each triangle stores three comlete vertices. If desired, we then
transform this to an indexed face set layout as follows: First, we
tagging each vertex with an int specifying both the triangle it
belongs to, and which of the triangle's three vertices it is. We then
sort this array of vertices by vertex position (using
\code{thrust::sort}), which necessarily means that shared vertices
will then be in adjacent locations.

Next, we execute a CUDA kernel that for each vertex in that sorted
array, checks if that vertex has a predecessor at the same position,
and returns if this is the case; otherwise, it assigns a unique index
array position to this vertex by increasing another atomic counter,
and writes the vertex to the final vertex array, and the index to the
vertex's relevant posiiton in the final index array, if such arrays
have been provided.  As above, we use the same trick of using the same
kernel for both counting and final writing by simply passing null
output arrays in the first pass, then allocating these arrays based on
the atomic counter's value, and re-run a second time with the actually
allocated arrays.

Pseudo-code for this algorithm is given in
Algorithm~\ref{alg:meshing}).  Obviously, if one has an upper bound on
the number of triangles respectively vertices generated the first of
these passes can be omitted; as can be the entire generation of shared
vertex array if a fat triangle layout is sufficient.

\begin{figure}[h]
  \vspace{-2em}
  \relsize{-1}{
    \begin{lstlisting}
      // get input
  thrust::host_vector<Cells> input = readInput();
  
  // upload and sort cells for binary search
  // in cell location kernel
  thrust::device_vector<Cells> d_cells = input;
  thrust::sort(d_cells);
  
  // create empty vertex buffer and atomic counter
  thrust::device_vector<int> = d_atomic(1);

  // first pass to count tris: run extraction kernel
  // with empty output
  thrust::device_vector<FatVertex> d_fatVtx = empty;
  extractIsoSurfaces<<<8*d_cells.size()>>>
       (d_cells,d_atomic,d_fatVtx);
  
  // allocate output array, and rerun
  d_fatVtx.resize(3*d_atomic[0]); d_atomic[0]=0;
  extractIsoSurfaces<<<8*d_cells.size()>>>
       (d_cells,d_atomic,d_fatVtx)
  
  // now have 'fat' fatVtx, convert to triangle mesh:
  // count fatVtx with dummy output arrays
  d_atomic[0] = 0;
  createIndexedFaces<<<d_fatVtx.size()>>>
       (d_fatVtx,d_atomic,null,null)
  
  // allocate output arrays, and rerun
  thrust::device_vector<float3> d_vtx(d_atomic[0]);
  thrust::device_vector<int3> d_idx(d_fatVtx.size()/3);
  d_atomic[0] = 0;
  createIdxedFaces<<<d_fatVtx.size()>>>
      (d_fatVtx,d_atomic,d_vtx,d_idx)

  // download to host, and done
  thrust::host_vector<float3> h_vtx = d_vtx;
  thrust::host_vector<int3> h_idx = d_idx;
  \end{lstlisting}
  }\vspace{-1em}
  \caption{\label{fig:meshing}
    \label{alg:meshing}
    Thrust-based extraction of iso-surface,
    and generation of shared vertex array. Our CUDA kernels for
    extracting the iso-surface and creating the indexed face list are
    described in the text.
  \vspace*{-2em}}
\end{figure}

\section{Results}

To both validate and evaluate our algorithm we have implemented a
variant using both CUDA and thrust (an earlier CPU variant used
\code{std::sort} for sorting, \code{std::lower\_bound} for binary
search, and \code{tbb} for parallelization), and ran it on a
workstation with a Intel i7-7820x CPU (8 cores, 3.6GHz), 64~GBs or
RAM, SSDs, and a NVIDIA RTX~8000 GPU.

The models we used for testing are given in Figure~\ref{fig:models},
and span a wide range in both model complexity and formats. In
particular, we point out that though Astro, Cloud, and Landing Gear
have \emph{some} sort of input mesh structure, the \emph{Exajet} and
\emph{impact} data sets come as unsorted lists of individual AMR cells
to which previous level-grid based approaches could not immediately be
applied. Exajet also comes with ``holes'' in the sense that parts
inside the airplane are not covered by any cells; using our method
these will automatically be detected as model boundaries (even though
some are \emph{inside} the model), and handled correctly without any
special effort.

As can be seen in the timings in Figure~\ref{fig:models} our algorithm
is not real-time, but still pretty fast, taking only a few seconds
even for complex data sets with hundreds of millions of cells, and
hundreds of millions of generated triangles.

\section{Summary and Conclusion}

In this paper, we have described an intriguingly simple technique for
generating the dual mesh of an arbitrary cell-centered AMR data set
that, though developed independently, may best be viewed as a simpler
and somewhat more general re-formulation of the epsilon-box snapping
method by Moran et al. In particular, the technique operates by
``snapping'' the right set of logical dual cells to actual AMR cells
using a fast cell location kernel that itself operates solely on an
array of sorted input cell coordinates. The method is intriguingly
simple to implement, does not make any assumptions on input type or
hierarchy structure, parallelizes trivially, lends itself well to
efficient implementation on both CPU and GPU, and performs well even
for highly non-trivial inputs.

Using this technique for generating dual cells we then demonstrated
its application to the easy, fast, and parallel extraction of
crack-free iso-surfaces; resulting in a framework that, using a CUDA
implementation, can generate the iso-surface of even complex AMR data
sets in seconds. Arguably the most defining trait of our approach is
its simplicity that doesn't requiring complex case tables, and works
in all cases including data sets with ``holes'', multiple level
boundaries, etc.  This simplicity not only facilitates simple code
when implementing it, it also means that the algorithm is naturally
parallel, and thus maps well to both multi-core CPUs and GPUs.

\paragraph*{Future work.}
In future, we would be very interested in seeing a implementation of
this algorithm within VTK, and, in particular, within
VTK-m~\cite{vtk-m} (to which the algorithms parallel nature should
lend very well), and to have it be used by actual end users of AMR
software.  Also, it would be interesting to see if our method could be
integrated into an AMR volume ray tracing framework to generate an
interpolant in the same way as originally done by Moran et al.

\section*{Acknowledgements}

The Exajet data set is courtesy Pat Moran, NASA; the Landing Gear is
courtesy Mike Barad (and others), NASA.  \textbf{TAC Molecular Cloud
  (``Cloud'')} and \textbf{Princeton Stellar Cluster Wind (``Wind'')}
are astrophysics simulations from the Theoretical Astrophysics Group
in Cologne, and Princeton, respectively.  \textbf{LANL Deep Water
  Impact (``Impact'')} is a simulation of an Asteroid Ocean Impact
computed with xRage, and courtesy of Patchett et al.

We would also like to express our gratitude to Will Usher, Nate
Morrical, Nate Marshak, Stefan Zellmann, Dave deMarle, Pat Moran, and
Gunther Weber, for either help and/or fruitful discussions/feedback
while working on this project: It is greatly appreciated!

\footnotesize
\bibliographystyle{eg-alpha}
%\bibliographystyle{abbrv-doi}
%\bibliographystyle{ACM-Reference-Format}
%\bibliography{sample-bibliography}
\bibliography{dualmesh}

\newpage
\section*{Appendix}

\subsection*{Accompanying Sample Code}

Sample CUDA code for the method described in this paper
has been made available via github,
on \code{https://github.com/ingowald/cudaAmrIsoSurfaceExtraction}.

\end{document}